\documentclass[a4paper,11pt]{article}
\usepackage{float}
\usepackage{jheppub}

\usepackage{ifpdf}
\usepackage{xcolor}
\usepackage{amsmath,amssymb,amsfonts,bm,amscd}
\usepackage{amscd}
\usepackage{mathtools}
\usepackage{graphicx, wrapfig}
\usepackage{multirow}
\usepackage{verbatim}
\usepackage{appendix}
\usepackage{fancybox}
\usepackage{slashed}
\usepackage{extarrows}
\usepackage{braket}
\usepackage{url}

\usepackage{bbold}

\renewcommand{\d}{\text{d}}
\newcommand{\bZ}{\mathbb{Z}}
\newcommand{\bC}{\mathbb{C}}
\newcommand{\bN}{\mathbb{N}}
\renewcommand{\d}{\text{d}}
\newcommand{\id}{\text{id}}

\newcommand\bI{\mathbb I}
\newcommand\coker{\text{coker}}

\usepackage{tikz}
\usetikzlibrary{decorations.markings}
\usetikzlibrary{decorations.pathreplacing,angles,quotes}
\usetikzlibrary{decorations.pathmorphing,shapes}
\newcounter{sarrow}

\definecolor{WScolor}{RGB}{191,191,255}
\definecolor{WScolor light}{RGB}{224,224,255}
\definecolor{DefectColor}{RGB}{0,0,128}
\tikzset{
	defect/.style={color=DefectColor, line width=1.5},
	arrow position/.style={postaction={decorate,decoration={
		markings,
		mark=at position #1 with {\arrow{>}}
	}}},
	opp arrow position/.style={postaction={decorate,decoration={
		markings,
		mark=at position #1 with {\arrow{<}}
	}}},
	defect node/.style={circle,inner sep=1.5pt,fill, DefectColor},
	snake it/.style={
		-stealth,
		decoration={
			snake, 
    		amplitude = .4mm,
    		segment length = 2mm,
    		post length=0.9mm
    	},
    	decorate
    }
}

\title{Phase transitions in GLSMs and defects}

\author[1]{Ilka Brunner,}
\author[2]{Fabian Klos,}
\author[3]{Daniel Roggenkamp}
\affiliation[1]{Arnold Sommerfeld Center, Ludwig-Maximilians-Universit\"at,\\
Theresienstra{\ss}e 37, 80333 M\"unchen, Germany}
\affiliation[2]{Institut f\"ur Theoretische Physik, Universit\"at Heidelberg,\\
Philosophenweg 19, 69120 Heidelberg, Germany}
\affiliation[3]{Institut f\"ur Mathematik, Universit\"at Mannheim,\\
B6, 26, 68131 Mannheim, Germany}

\abstract{In this paper, we construct defects (domain walls) that connect different phases of two-dimensional gauged linear sigma models (GLSMs), as well as defects that embed those phases into the GLSMs. Via their action on boundary conditions these defects give rise to functors between the  D-brane categories, which  respectively describe the transport of D-branes between different phases, and embed the D-brane categories of the phases into the category of D-branes of the GLSMs.}

\preprint{LMU-ASC 01/21}

\begin{document}
\cornersize{1}

\maketitle

\section{Introduction}
The topic of this paper are 
two-dimensional gauged linear sigma models with $U(1)$ gauge groups\footnote{Indeed, our discussion easily generalizes to models with arbitrary abelian gauge groups.}. 
These are 
2d $N=(2,2)$ supersymmetric gauge theories coupled to  chiral superfields carrying possibly different charges under the $U(1)$ gauge group, such that the respective superpotentials $W$ are $U(1)$ invariant.

As is well known, gauged linear sigma models exhibit different phases for different ranges of the 
Fayet-Iliopoulos parameter $r$ associated to the $U(1)$ gauge group \cite{Witten:1993yc}. For non-anomalous gauged linear sigma models, where axial and vector $R$-symmetries are preserved at the quantum level, the RG flow drives the GLSM to a (K\"ahler) moduli space of superconformal field theories parametrized by the complexified Fayet-Iliopoulos parameter $t$. The phases correspond to 
different domains of this moduli space.
In contrast, in the anomalous case, the FI parameter is a running coupling constant, and 
the different phases correspond to fixed points under the RG flow.

The phases typically exhibit gauge symmetry breaking. For instance, in geometric phases, in which the theory can be effectively described by a non-linear sigma model, the gauge group is typically completely broken. 
On the other hand,  in phases in which the theory can be described by Landau-Ginzburg models (Landau-Ginzburg phases),
a finite subgroup of the gauge group remains unbroken and survives as an orbifold group.

The question we will address in this paper is how the boundary sectors  (i.e. the D-branes) behave under transitions between different phases of GLSMs. 

The transport of D-branes  between different  phases of abelian gauged linear sigma models has initially been studied in \cite{Herbst:2008jq} for the non-anomalous ``Calabi-Yau'' case. Results on the  anomalous ``non-Calabi-Yau'' case appeared more recently in \cite{Hori:2013ika,Clingempeel:2018iub}. With a careful analysis, the authors of these papers obtain a prescription of the D-brane transport on the level of individual D-branes:
Starting in one phase, a D-brane is first lifted to the gauged linear sigma model. This lift is a priori not unique, but requires certain choices. These choices correspond to the homotopy classes of paths along which the D-brane can be smoothly transported in parameter space (``grade restriction rule''). Having lifted a D-brane in such a way that it can be smoothly transported along the chosen path, one only has to push down the lift to the other phase.

In this paper, we want to revisit the transport of D-branes between different phases from an alternative point of view and in particular give a uniform and functorial description of it. 
Our basic idea is to construct  suitable  supersymmetry preserving defect lines (domain walls) that connect different phases of a GLSM (or a GLSM and one of its phases). Such defects can be inserted along lines in space-time which separate space-time domains in which different phases of a GLSM are realized. 
\begin{equation}\tikz[baseline=10]{
	\fill[WScolor light] (-2,0) rectangle (4,1);
	\draw[defect] (1,0) -- (1,1);
	\node at (1.55,.2) {$$};
	\node at (-.5,.5) {phase$_1$};
	\node at (2.5,.5) {phase$_2$};
}\;\tikz[baseline=-4]{
}\;\tikz[baseline=10]{
	\fill[WScolor ] (-2,0) rectangle (1,1);
	\fill[WScolor light] (1,0) rectangle (4,1);
	\draw[defect] (1,0) -- (1,1);
	\node at (-.5,.5) {GLSM};
	\node at (2.5,.5) {phase};
	\node at (1.3,.2) {$$};
}\end{equation}
Supersymmetry preserving defects can be merged with boundaries (``fusion'') and in this way give rise to an action on D-branes. 
\[\begin{aligned}\tikz[baseline=17,scale=0.8]{
		\fill[WScolor] (-1,.25) rectangle (2,1.25);
		\draw[defect] (2,.25) -- (2,1.25);
		\node at (1.7,.45) {$B$};
		\node at (.5,.75) {GLSM};
	}\quad&\longmapsto\quad\tikz[baseline=17, scale=0.8]{
		\fill[WScolor light] (-2,.25) rectangle (1,1.25);
		\fill[WScolor] (1,.25) rectangle (4,1.25);
		\draw[defect] (1,.25) -- (1,1.25);
		\draw[defect] (4,.25) -- (4,1.25);
		\node at (.7,.45) {$D$};
		\node at (-.5,.75) {phase};
		\node at (2.5,.75) {GLSM};
		\node at (3.7,.45) {$B$};
	}\phantom{*}&\tikz[baseline=-2]{
	\path (0,0) edge[snake it] (0.7,0);
}\;\phantom{**}
	\tikz[baseline=17,scale=0.8]{
		\fill[WScolor light] (-1,.25) rectangle (2.5,1.25);
		\draw[defect] (2.5,.25) -- (2.5,1.25);
		\node at (1.7,.45) {$D\otimes B$};
		\node at (0.4,.75) {phase};
	}
\end{aligned} \]
This action is functorial, and hence any supersymmetric defect between two theories yields a 
functor between the respective D-brane categories.

In order to find the defects describing the D-brane transport in GLSMs, we will follow a strategy analogous to the one employed in \cite{Brunner:2007ur} to describe the behavior of D-branes under RG flows. The basic idea  of that paper is to consider an RG flow of a UV theory triggered by a perturbation that is restricted to a domain of space-time. \begin{equation}\label{pic:flow} \tikz[baseline=10]{
	\fill[WScolor] (-2,0) rectangle (4,1);
	\draw[densely dashed] (1,0) -- (1,1);
	\node at (-.5,.5) {perturbed UV};
	\node at (2.5,.5) {UV};
}\;\tikz[baseline=-4]{
\node at (0.7,0.4) {RG flow};
	\path (0,0) edge[snake it] (1.5,0);
}\;\tikz[baseline=10]{
	\fill[WScolor light] (-2,0) rectangle (1,1);
	\fill[WScolor] (1,0) rectangle (4,1);
	\draw[defect] (1,0) -- (1,1);
	\node at (-.5,.5) {IR};
	\node at (2.5,.5) {UV};
	\node at (1.3,.2) {$R$};
}\end{equation}
The RG flow drives the theory to the IR inside this domain, and leaves it at the UV fixed point outside of it, thereby creating a defect line on the domain boundary, which separates the UV and IR theories. This kind of construction has found applications in a variety of contexts, for supersymmetric as well as non-supersymmetric theories, for marginal as well as relevant perturbations, see \cite{Bachas:2007td,Konechny:2015qla,Gaiotto:2012np}. This ``RG defect'' encodes
the behavior of the boundary sector under the flow. In case that the models are $N=(2,2)$ supersymmetric and that the respective perturbations preserve half of the supersymmetry, the resulting defects are also supersymmetric. Their fusion with supersymmetric boundary conditions can easily be calculated, and the respecting functors can be determined.

Supersymmetric RG defects have some important characteristic properties that were discussed in \cite{Klos:2019axh}. For example, contraction of small unperturbed UV regions inside IR patches does not change correlation functions. This can be interpreted in terms of fusion, and one obtains the identity
 \[\tikz[baseline=20,scale=0.7]{
		\fill[WScolor light] (0,0) rectangle (2,1.5);
		\fill[WScolor] (2,0) rectangle (4,1.5);
		\fill[WScolor light] (4,0) rectangle (6,1.5);
		\draw[defect] (2,0) -- (2,1.5);
		\draw[defect] (4,0) -- (4,1.5);
		\node at (3,.75) {UV};
		\node at (1,.75) {IR};
		\node at (5,.75) {IR};
		\node at (2.3,.2) {R};
		\node at (4.3,.2) {T};
	}\tikz[baseline=6]{
	\path (0,0) edge[snake it] (0.7,0);
}
\tikz[baseline=20,scale=0.7]{
		\fill[WScolor light] (-1,0) rectangle (2,1.5);
		\fill[WScolor light] (2,0) rectangle (5,1.5);
		\draw[defect] (2,0) -- (2,1.5);
		\node at (3.5,.75) {$IR$};
		\node at (0.5,.75) {$IR$};
				\node at (2.8,.2) {$R\otimes T$};
	}
	\tikz[baseline=6]{
	\node at (0,0) {$\cong$};}
	\tikz[baseline=20,scale=0.7]{
		\fill[WScolor light] (-1,0) rectangle (2,1.5);
		\fill[WScolor light] (2,0) rectangle (5,1.5);
		\draw[densely dashed] (2,0) -- (2,1.5);
		\node at (3.5,.75) {$IR$};
		\node at (0.5,.75) {$IR$};
				\node at (2.5,.3) {$I^{\text{IR}}$};
	}
\]
where $R$ is the RG defect, 
$T$ is an adjoint of it and $I^{\text{IR}}$ is the invisible identity defect of the IR theory.
Thus, fusion $R\otimes T\cong I^{\rm IR}$ yields  the identity defect in the IR. This also implies that the fusion of $R$ and $T$ in the opposite order gives rise to a projection defect $P=T\otimes R$ in the UV theory. This defect is idempotent with respect to fusion and can be used to realize the IR theory inside the UV theory.

In this paper we will explain how similar ideas can be employed in the context of abelian GLSMs to obtain transition defects between different phases of GLSMs as well as defects between GLSMs and their phases. In this way, we derive a novel method for  brane transport  and in particular recover the grade restriction rule of  \cite{Herbst:2008jq,Hori:2013ika,Clingempeel:2018iub} from this point of view. While we do have compatible results, our derivations are rather different from those of \cite{Herbst:2008jq,Hori:2013ika,Clingempeel:2018iub}. In our discussion we decouple all gauge degrees of freedom and   merely take into account the matter sector, the only remnant of the gauge symmetry being an equivariance condition. This subsector is under good control and still captures the physics of the (B-type) supersymmetry preserving sector, including perturbations, boundary conditions and defects. Our arguments mainly rely on the rigidity of defect constructions in this setting. The defects we construct on this level directly mediate between  the different phases and do not exhibit an explicit $t$ dependence.

In section~\ref{sec:transition} we will outline the general ideas.
In particular, we will discuss the construction of defects connecting different phases of a GLSM and explain how they factorize into defects lifting phases to the GLSM and those pushing down the GLSM to phases.
Furthermore, we will introduce projection defects which realize the phases inside the GLSM. Their 
action on the D-brane category of the GLSM corresponds, in the language of 
\cite{Herbst:2008jq,Clingempeel:2018iub}, to the projection of GLSM branes to 
``grade restricted" representatives. 

The starting point for the construction of the transition defects is the identity defect of the GLSM theory which will be constructed explicitly 
in section~\ref{sec:identity}. For this, one needs to generalize the known constructions of orbifold identity defects \cite{Brunner:2007ur,Carqueville:2012dk}  to continuous (abelian) groups. We show that this can be achieved (in the context of equivariant matrix factorizations) by introducing  new bosonic fields constrained to the defect. Our expectation is that this idea can be applied more generally in topological field theories.

In section~\ref{sec:example}, we will 
illustrate our general construction in an explicit class of examples, namely the 
$U(1)$-gauged linear sigma models with two chiral fields, $X$ and $P$, and superpotential 
$W=P^{d'}X^d$, $d'<d$. These anomalous models have two different Landau-Ginzburg orbifold phases. 
The UV phase is described by the Landau-Ginzburg orbifold with superpotential $W= X^d$ and orbifold group ${\mathbb Z}_d$, and the IR phase by the Landau-Ginzburg orbifold with superpotential 
 $W=P^{d'} $ and orbifold group ${\mathbb Z}_{d'}$.
 Along the RG flow, $d-d'$ vacua decouple to a Coulomb branch taking with them a set of D-branes. All of this is encoded in the transition defects we construct here. We recover various results from \cite{Herbst:2008jq,Hori:2013ika,Clingempeel:2018iub} on brane transport. Moreover, in this example
the phase transition between UV and IR phase of the GLSM corresponds to a well understood RG flow between the 
Landau-Ginzburg models describing the UV and IR phases \cite{Dabholkar:2001wn}. RG defects for these flows have been constructed in \cite{Brunner:2007ur} and we indeed find that  our transition defects between UV and IR phase  agree with the respective RG defects.

\section{Phases of GLSMs and  defects}\label{sec:transition}

\subsection{Phases of GLSMs}

We are considering two-dimensional  $\mathcal{N}=(2,2)$ gauged linear sigma models with  abelian gauge groups \cite{Witten:1993yc}. By  $X_i$, $i=1,\ldots, n$ we denote the chiral superfields of the theory. Their representation under the gauge group $U(1)^k$ is specified by the charge matrix $Q_i^a$, where $i=1,\ldots, n$ and $a=1,\ldots, k$. For each $U(1)$-factor  of the  gauge group  the theory contains  a field strength multiplet, a twisted chiral field $\Sigma_a$,  $a=1,\ldots,k$. 
We also allow for a superpotential $W$, which is a gauge invariant polynomial in the superfields $X_i$. 

The classical bosonic potential for the scalar parts $x_i$ of the chiral superfields $X_i$ and $\sigma_a$ of the  twisted chiral fields $\Sigma_a$ is given by 
\begin{equation}\label{eq:general scaler potential}
	U = \sum_{i=1}^n \left\vert \sum_{a=1}^k Q^a_i \sigma_a x_i\right\vert^2 + \frac{e^2}2 \sum_{a=1}^k \left( \sum_{i=1}^n Q^a_i \vert x_i\vert^2 - r^a  \right)^2 + \sum_{i=1}^n \left\vert\frac{\partial W}{\partial x_i}(x_1, ..., x_n)\right\vert^2.
\end{equation}
Here, $r^a\in\mathbb{R}$ is the Fayet-Iliopoulos (FI) parameter of the $a$th $U(1)$ gauge factor. Together with the corresponding $\theta$-angle $\theta^a$ it forms a complex parameter $t^a=r^a-i\theta^a$. 
(The gauge couplings, $e$ of the $U(1)$-factors are assumed to be equal.)

The classical vacuum manifold is obtained as the space of solutions to the equation $U=0$ modulo gauge transformations. Its nature depends crucially on the specific values of $(r^1, ..., r^k)$. The subspace parametrized by the expectation values of the matter fields is commonly referred to as the Higgs branch, whereas the scalars $\sigma_a$ parametrize a Coulomb branch.  Phases in which the gauge group is completely broken   and all modes transverse to $\{U=0\}$ are massive are called \emph{geometric phases}. In these phases, the Higgs branch is effectively described by a non-linear  sigma model with target space
$\{U=0\}/U(1)^k$. If on  the other hand the space of vacua $\{U=0\}/U(1)^k$ consists of a single point and all modes transverse to the orbit of the complexified gauge group remain massless, the Higgs branch is effectively described by a Landau-Ginzburg (orbifold) theory. Such phases are called \emph{Landau-Ginzburg phases}. Besides these extreme ones, GLSMs can also exhibit various mixed phases.  Furthermore, classically at  $r=0$, all fields can be $0$. This means that some of the $\sigma_a$ can be non-zero, and parametrize vacua on another branch, the Coulomb branch. 

An important quantum effect is the renormalization of the Fayet-Iliopoulus parameters: 
\begin{equation}\label{eq:running}
 r^a(\mu)= r^a_{UV}  + Q^a_{tot} \log\frac{\mu}{M_{UV}}. 
\end{equation}
Here $M_{UV}$ denotes a $UV$ energy scale, $\mu$ the scale under consideration, and $Q^a_{tot}=\sum_{i=1}^NQ^a_i$ is the total charge of the respective $U(1)$ factor. 
If $Q^a_{tot}=0$ for all $a$, the axial R-symmetry of the theory is non-anomalous and the FI parameters do not run. The $t^a$ are genuine parameters of the theory. This case is called the ``Calabi-Yau case''. 

If one of the total charges is non-zero, the respective FI parameter does run under the RG flow. The direction of the running and with it the nature of the low energy IR phase is determined by the sign of the total charge.

In general, the low energy IR phase to which the system is driven by the RG flow consists of several branches. In the specific example considered in section~\ref{sec:example}, there is a Higgs branch described by a Landau-Ginzburg model as well as several massive vacua located on a Coulomb branch. 

Note, that also in the anomalous case the system can explore various different phases \cite{Clingempeel:2018iub}. For this, one chooses $r_{UV}$ such that  at some intermediate energy scale the system is  well described by the desired phase. Our main example in section~\ref{sec:example} features, besides the IR phase an additional phase corresponding to the UV fixed point. This UV phase is a Landau-Ginzburg phase as well, but in contrast to the IR phase, it is a pure Higgs phase, i.e.~it does not have additional Coulomb vacua.

\subsection{Phases of GLSMs and defects}

We now want to obtain defects describing the transition between different phases of the same GLSM, much in line with the construction of RG defects reviewed in the introduction. The general idea is to start with the identity or invisible defect $I^\text{GLSM}$  in the GLSM, and to push the GLSM down to different phases on the two sides of it:
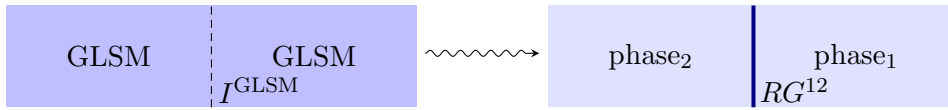
\begin{equation}\nonumber \tikz[baseline=10,scale=0.9]{
	\fill[WScolor] (-2,-.25) rectangle (4,1.25);
	\draw[densely dashed] (1,-.25) -- (1,1.25);
	\node at (1.7,0) {$I^\text{GLSM}$};
	\node at (-.5,.5) {GLSM};
	\node at (2.5,.5) {GLSM};
}\;\tikz[baseline=-4]{
	\path (0,0) edge[snake it] (1.5,0);
}\;\tikz[baseline=10,scale=0.9]{
	\fill[WScolor light] (-2,-.25) rectangle (1,1.25);
	\fill[WScolor light] (1,-.25) rectangle (4,1.25);
	\draw[defect] (1,-.25) -- (1,1.25);
	\node at (-.5,.5) {phase$_2$};
	\node at (2.5,.5) {phase$_1$};
	\node at (1.6,0) {$RG^{12}$};
}\end{equation}
A priori this requires tuning the $t^a$ on the two sides of the defect to different regimes. We avoid doing this explicitly by going to an extreme UV limit of the theory, in which the gauge coupling $e$ becomes very small
and the gauge sector decouples \cite{Herbst:2008jq}. In this limit, the theory reduces to the matter sector, describing the Higgs branch of the original theory. 
The gauge group still acts on the matter fields, and physical observables must be gauge invariant. 
The defects are B-type supersymmetric, and depend on the parameters $t^a$ only indirectly through stability conditions that can  be added consistently to this sector.
In this setup, the transition to a phase restricts  the allowed field configurations and breaks the (remnant of the) gauge symmetry to a subgroup. The details strongly depend on the respective phase. 

For instance, the class of examples we will discuss in section~\ref{sec:example} features Landau-Ginzburg phases. In such phases, some of the fields obtain a vacuum expectation value, reducing the spectrum of massless excitations and breaking the gauge symmetry to a finite subgroup. Pushing down to such a phase then involves setting the respective fields to their vacuum expectation values and relaxing the invariance condition accordingly.
The general strategy outlined here should be applicable to any phases of abelian GLSMs. Transitions to geometric phases will be discussed in a forthcoming paper \cite{forthcoming}.

Note that one obtains a possibly different defect for every homotopy class of paths
connecting two given phases 
 in the parameter space spanned by the $t^a$.
Thus, in general there will not be one transition defect descending from the gauged linear sigma model, but many, and the choices of defects should correspond to choices of paths. On the other hand, there can be more RG flows and with it RG defects between different phases of a GLSM then the ones described within the GLSM.

Indeed, the transition defects $RG^{12}$ between two phases of a GLSM factorize over the GLSM, i.e.~
$RG^{12}$ can be obtained as the fusion of a defect $T^1$ from phase$_1$ to the GLSM and a defect $R^2$ from the GLSM to phase$_2$:
$$
RG^{12}\cong R^2\otimes T^1 \ .
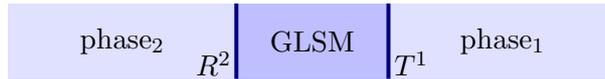
$$
\[\tikz[baseline=10]{
	\fill[WScolor light] (-2,0) rectangle (1,1);
	\fill[WScolor] (1,0) rectangle (3,1);
	\fill[WScolor light] (3,0) rectangle (6,1);
	\draw[defect] (1,0) -- (1,1);
	\draw[defect] (3,0) -- (3,1);
	\node at (4.5,.5) {phase$_1$};
	\node at (2,.5) {GLSM};
	\node at (-.5,.5) {$\text{phase$_2$}$};
	\node at (.7,.2) {$R^2$};
	\node at (3.3,.2) {$T^1$};
}\]
The defects $T^1$ and $R^2$ are obtained from the GLSM identity defect, by pushing down only on one side. $T^1$ is obtained by pushing down $I^\text{GLSM}$ on the right to phase$_1$ and $R^2$ by pushing down $I^\text{GLSM}$ on the left to phase$_2$:
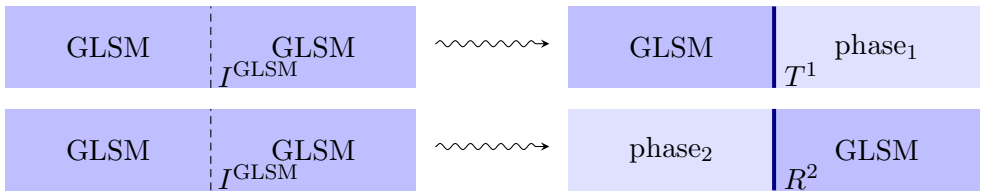
\begin{eqnarray}\nonumber \tikz[baseline=10,scale=0.9]{
	\fill[WScolor] (-2,-0.1) rectangle (4,1.1);
	\draw[densely dashed] (1,-0.1) -- (1,1.1);
	\node at (1.7,.1) {$I^\text{GLSM}$};
	\node at (-.5,.5) {GLSM};
	\node at (2.5,.5) {GLSM};
}\;&\tikz[baseline=-4]{
	\path (0,0) edge[snake it] (1.5,0);
}&\;\tikz[baseline=10,scale=0.9]{
	\fill[WScolor ] (-2,-.1) rectangle (1,1.1);
	\fill[WScolor light] (1,-.1) rectangle (4,1.1);
	\draw[defect] (1,-.1) -- (1,1.1);
	\node at (-.5,.5) {GLSM$$};
	\node at (2.5,.5) {phase$_1$};
	\node at (1.4,.1) {$T^1$};
}
\\
\nonumber \tikz[baseline=10,scale=0.9]{
	\fill[WScolor] (-2,-.1) rectangle (4,1.1);
	\draw[densely dashed] (1,-.1) -- (1,1.1);
	\node at (1.7,.1) {$I^\text{GLSM}$};
	\node at (-.5,.5) {GLSM};
	\node at (2.5,.5) {GLSM};
}\;&\tikz[baseline=-4]{
	\path (0,0) edge[snake it] (1.5,0);
}&\;\tikz[baseline=10,scale=0.9]{
	\fill[WScolor light] (-2,-.1) rectangle (1,1.1);
	\fill[WScolor ] (1,-.1) rectangle (4,1.1);
	\draw[defect] (1,-.1) -- (1,1.1);
	\node at (-.5,.5) {phase$_2$};
	\node at (2.5,.5) {GLSM$$};
	\node at (1.4,.1) {$R^2$};
}
\end{eqnarray}
The $R^i$ encode the push down from the GLSM to phase$_i$ and the $T^i$, the embedding of phase$_i$ into the GLSM. The functors associated to those defects 
describe the respective operation on D-brane categories. 

Note that in the same way that there can be several transition defects $RG^{12}$ between different phases
we expect  more than one possible defect $T^i$ lifting the phase to the GLSM.\footnote{Indeed, one could also push the path dependence on the $R^j$.} This will be discussed in more detail for the concrete example in section~\ref{sec:example}.

Since the transition between one and the same phase has to be trivial, the defects $R^i$ and $T^i$ have to satisfy the condition
\begin{equation}\label{eq:RT}
R^i \otimes T^i\cong
I^{{\text{phase}}_i},
\end{equation}
where $I^{{\text{phase}}_i}$ is the invisible defect of phase$_i$. This implies that the combination
$$P^{i}=T^i \otimes R^i$$
is a projection defect from the GLSM to itself. That means $P^i$ is an idempotent with respect to fusion,
$P^{i}\otimes P^i\cong P^i$, and realizes phase$_i$ inside the GLSM in the sense of \cite{Klos:2019axh}. In particular,
the corresponding functor projects the category of D-branes of the GLSM onto the image of the functor associated to $T^i$. Thus, the phase$_i$ branes are realized by $P^i$-invariant branes in the GLSM. 
Indeed, the latter
precisely play the role of the branes called {\emph{grade restricted}} in \cite{Herbst:2008jq} and the action of the projection defects corresponds to the operation of associating to a GLSM brane a grade restricted representative. 
We will see this explicitly in the example discussed in section~\ref{sec:example}.

We have collected the various defects and their actions on D-branes in the following diagram:
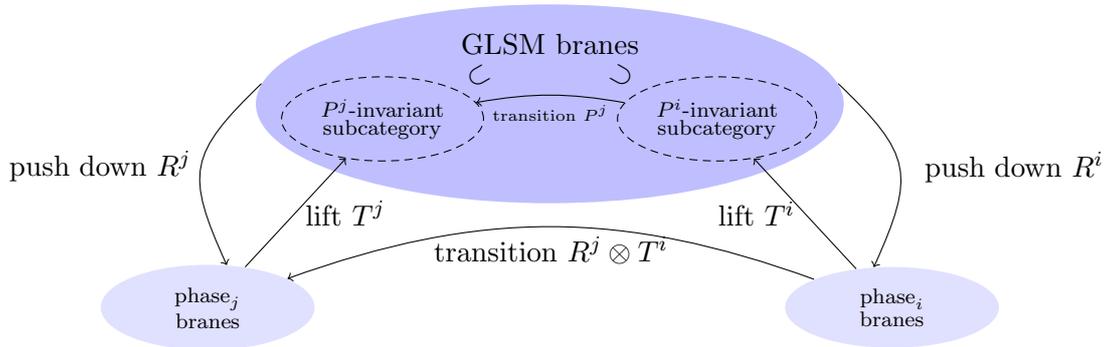
\[\tikz{
\node[ellipse, fill=WScolor,minimum width=220,minimum height=75](A) at (0,-.8){};
	\node (F) at (0,0) {GLSM branes};
	\node[ellipse, fill=WScolor, draw, style=densely dashed] (B) at (-2.2,-1) {$\substack{P^j\text{-invariant}\\\text{subcategory}}$};
	\node[ellipse, fill=WScolor, draw, style=densely dashed] (C) at (2.2,-1) {$\substack{P^i\text{-invariant}\\\text{subcategory}}$};
	\node[ellipse, fill=WScolor light,minimum width =80] (D) at (-4.5,-3.5) {$\substack{\text{phase}_j\\\text{branes}}$};
	\node[ellipse, fill=WScolor light,minimum width =80] (E) at (4.5,-3.5) {$\substack{\text{phase}_i\\\text{branes}}$};
	
	\path (A) edge[bend left=35,looseness=1.4,->] node[right]{\phantom{*}push down $R^i$} (E);
	\path (A) edge[bend right=35,looseness=1.4,->] node[left]{push down $R^j$} (D);
	\path (B) edge[<-, bend left=10] node[below]{\tiny transition $P^j$} (C);
	\path (D) edge[->] node[right]{lift $T^j$} (B);
	\path (E) edge[->] node[left]{lift $T^i$} (C);
	\path (D) edge[<-, bend left=20] node[below]{transition $R^j\otimes T^i$} (E);
	\path (B) -- node[rotate=30]{$\subset$} (F);
	\path (C) -- node[rotate=150]{$\subset$} (F);
}\]
Note that along an RG flow, Higgs vacua can migrate to the Coulomb branch and become massive. Since we only include the Higgs branch in our discussion, we cannot see this directly. Instead, we observe that D-branes attached to those vacua decouple  from the theory.
This decoupling of D-branes is encoded in the defects introduced above. They can be constructed out of the identity defect of the respective GLSM, which will be introduced in the next section.

\subsection{GLSM Identity Defects}\label{sec:defects}

Starting point of our construction are the identity defects of abelian GLSMs, which have not appeared in the literature so far. As discussed above, we will focus on the Higgs branch of the respective GLSM. In particular, we will decouple the gauge sector and only consider the $U(1)^k$-orbifold of the matter sector. 
The relevant defects and D-branes can then be described by means of $U(1)^k$-equivariant matrix factorizations.

Before discussing the identity defects in GLSMs, we will briefly introduce some useful facts about matrix factorizations and discuss the identity defects in Landau-Ginzburg models. 

\subsubsection{Matrix factorizations and associated modules}\label{sec:mfmodules}

Defects in Landau-Ginzburg (orbifold) models are described by (equivariant) matrix factorizations of the superpotential, see the appendix for a brief discussion and references to the literature.
Consider a  matrix factorization 
\begin{equation*}
P:P_1
\\ 
	\tikz[baseline=0]{
		\node at (0,.5) {$p_1$};
		\draw[arrow position = 1] (-1.5,.2) -- (1.5,.2);
		\draw[arrow position = 1] (1.5,0) -- (-1.5,0);
		\node at (0,-.4) {$p_0$};
	}\, P_0 \,
\end{equation*}
of a polynomial $W$ in a polynomial ring $S=\bC[X_1,\ldots,X_n]$. Here,   $P=P_0\oplus P_1$ is a $\bZ_2$-graded free module  over $S$ and 
\[
	\d_P=\begin{pmatrix}0 & p_1 \\ p_0 & 0\end{pmatrix}\quad
\]
is an odd endomorphism of $P$ such that $\d_P^2 = W\cdot \id_P$.
In the following, instead of dealing with matrix factorizations we will often consider certain associated modules.
 To $P$ as above, one can associate the module 
\begin{equation*}
M_P=\text{coker}(p_1:P_1\otimes_SC\rightarrow P_0\otimes_SC)
\end{equation*}
over the respective quotient ring $C=S/(W)$. This module has a free two-periodic resolution defined by the matrix factorization \cite{eisenbud}
\begin{equation*}
\ldots\stackrel{p_0}{\longrightarrow} P_1\otimes_SC\stackrel{p_1}{\longrightarrow} P_0\otimes_SC\stackrel{p_0}{\longrightarrow} P_1\otimes_SC\stackrel{p_1}{\longrightarrow} P_0\otimes_SC\longrightarrow M_P\rightarrow 0 \ .
\end{equation*} 
Isomorphisms between modules $M_P$ and $M_Q$ associated to matrix factorizations $P$ and $Q$ of the same polynomial $W$ lift to these resolutions and give rise to isomorphisms of the respective matrix factorizations. 
We will make excessive use of this fact in our analysis. Indeed, the above argument works in the same way for any $C$-module $M_P$, which 
has a free resolution turning, after finitely many steps, into a two-periodic resolutions defined by the matrix factorization $P$. This point of view carries over to the case of equivariant matrix factorizations, c.f.~the appendix. 

\subsubsection{The identity defect in Landau-Ginzburg models}\label{sec:lgid}

The identity or invisible defect $I^W$ in a Landau-Ginzburg model with superpotential $W\in\bC[X_1,\ldots, X_n]$  is given by the following Koszul matrix factorization $(I,d_I)$ \cite{Kapustin:2004df}. 
Denote by $S_{(X)(X')}=\bC[X_1,\ldots,X_n,X_1',\ldots,X_n']$ the polynomial ring of the chiral fields on both sides of the defect. For later use, we will also denote by $S_{(X)(-)}=\bC[X_1,\ldots,X_n]$ the ring of chiral fields on the left and by $S_{(-)(X')}=\bC[X_1',\ldots,X_n']$ the ring of chiral fields on the right of the defect.
Then 
\begin{equation} \label{module}
	I=I_0\oplus I_1= {S}_{(X)(X')} \otimes \Lambda \left(V\right)
\end{equation}
is the tensor product of the algebra of chiral fields ${S}_{(X)(X')}$ with the exterior algebra of a 
vector space $V=\text{span}_\bC\{\theta_1,\ldots,\theta_n\}$ spanned by additional variables
$\theta_1,\ldots,\theta_n$. The latter correspond to fermions on the defect, and the $\bZ_2$-degree is inherited from the $\bZ$-degree of the exterior algebra.
The differential can be written as 
\begin{equation}
 \label{eq:LGid}
\begin{aligned} 
	\d_I &= \sum_{i=1}^n\left[ (X_i-X_i')\cdot \theta^*_i + \partial^{X,X'}_{i}W\cdot \theta_i \right] \\
	\text{with }\partial^{X,X'}_iW &= \frac{W(X_1', ..., X_{i-1}', X_i, ..., X_n) - W(X_1', ..., X_{i}', X_{i+1}, ..., X_n)}{X_i-X_i'}
\end{aligned}\end{equation}
where 
$\{\theta_1^*,\ldots,\theta_n^*\}$ denotes the dual to the basis $\{\theta_1,\ldots,\theta_n\}$, i.e.~$\theta_i^*(\theta_j )=\delta_{ij}$. 
Physically, the $\theta_i$ correspond to boundary fermions. Note that $\partial^{X,X'}_iW$ is a polynomial in the $X_i$ and $X_i'$.

This defect is the unit under fusion with other defects or boundaries, which means that for any defect $P$, the fusion $I^W \otimes P$ and $P \otimes I^W$ are isomorphic to $P$. This can be easily seen using the associated modules. To the identity defect, we can associate the ${C}_{(X)(X')}={S}_{(X)(X')}/(W(X_1,\ldots,X_n)-W(X_1',\ldots,X_n'))$-module
\[
M_I={C_{(X)(X')}}/(X_1-X_1', \dots, X_n-X_n') \ .
\]
This module has a free resolution, which after finitely many steps turns into the two-periodic resolution defined by the matrix factorization (\ref{eq:LGid}), see e.g. \cite{Enger:2005jk}.

Let $P'$ be a matrix factorization of $W(X_1',\ldots,X_n')$. Fusion of $I$ with $P'$ is given by the tensor product of the respective matrix factorizations $I\otimes P'$ \cite{Brunner:2007qu}. Let 
\begin{eqnarray}
C_{(-)(X')}&=&S_{(-)(X')}/(W(X_1',\ldots,X_n'))\,,\nonumber\\
\text{and}\;
C_{(X)(-)}&=&S_{(X)(-)}/(W(X_1,\ldots,X_n))\,.\nonumber
\end{eqnarray}
Then the $C_{(-)(X')}$-module 
$$
M_{P'}=\coker(p'_1:P'_1\otimes_{{S}_{(-)(X')}} C_{(-)(X')}\rightarrow P'_0\otimes_{S_{(-)(X')}} C_{(-)(X')})
$$
is associated to the matrix factorization $P'$.
Now, the 
tensor product  $M_I\otimes_{S_{(-)(X')}}M_{P'}$ is a $C_{(X)(-)}$-module, which has a 
free resolution turning, after finitely many steps,  into the tensor product matrix factorization $I\otimes P'$ of $W(X_1,\ldots,X_n)$. Of course $M_I\otimes_{S_{(-)(X')}}M_{P'}$ is isomorphic to the cokernel of the matrix $p_1$, obtained by setting all entries $X_i'$ in $p_1'$ to $X_i$. 
After all, the relations $X_i-X_i'$ in $M_I$ just set $X_i'$ to $X_i$. This module has a free resolution given by the matrix factorization $P$, which is obtained by replacing all the $X_i'$ in $P'$ by $X_i$. 
Hence, fusion with $I$ maps matrix factorizations to equivalent ones.

In case the Landau-Ginzburg model has a symmetry group $G_W$ which linearly acts on the chiral fields, $X_i\mapsto g(X_i)$ for $g\in G_W$ such that $W(g(X_1),\ldots,g(X_n))=W(X_1,\ldots,X_n)$, one can construct defects ${}_g I$ that implement the symmetry operation. The matrix factorizations of these symmetry defects are built on the same module as the identity matrix factorization, but the differential is twisted by the symmetry
\cite{Brunner:2007qu}
\[\begin{aligned}
	\d_{{}_g I} &= \sum_{i=1}^n\left[ (X_i-g(X_i'))\cdot \theta^*_i + \partial^{X,X'}_{i}W (X, g(X')) \cdot \theta_i \right] \,.
\end{aligned}\]

In the Landau-Ginzburg phases appearing here, the gauge group is not completely broken, and its remnant survives as a finite abelian orbifold group, acting by multiplying the chiral fields $X_i$ by phases.

Defects in Landau-Ginzburg orbifolds can be described by matrix factorizations of the superpotential which are equivariant with respect to the action of the orbifold group, see the appendix for a brief discussion and references.

The identity defect in Landau-Ginzburg orbifolds can be constructed from the identity defect of the unorbifolded LG model (\ref{module}), (\ref{eq:LGid}) using the method of images (i.e. summing images under the orbifold group and specifying a representation of the stabilizer subgroup).
Let $G$ be a finite abelian orbifold group. The identity defect of the orbifold model can be obtained by summing over all 
symmetry defects ${}_gI^\text{non-orb}$ of the unorbifolded Landau-Ginzburg model associated to orbifold group elements
\cite{Brunner:2007ur}
\begin{equation}
	I^\text{orb} = \bigoplus_{g\in G} {}_gI^\text{non-orb} \ .
\end{equation}
Note here that one has to orbifold by $G\times G$, the product of the orbifold groups on the left and on the right of the defect, but that the diagonal subgroup acts as an isomorphism on the non-orbifolded identity defect. Hence only a non-diagonal copy of $G$, which we take to be the copy $G_r$ acting trivially on the left of the defect contributes to the sum above. 

The module on which the orbifolded identity matrix factorization is built is therefore a direct sum of $|G|$ copies of the module (\ref{module}) associated to the identity defect in the unorbifolded Landau-Ginzburg model. We can regard it as a tensor product of the module $I^\text{non-orb}$ with the regular representation $V_\text{reg}$ of the group $G_r$: 
$I^\text{orb}\cong I^\text{non-orb}\otimes V_\text{reg}$.

The differential acts diagonally in the standard basis $g\in G_r$ of the regular representation,
while the orbifold group acts in this basis 
by permuting the copies of the modules $I^\text{non-orb}$ according to the group law. 

Since $G_r$ is finite and abelian, we can diagonalize the group action on $I^\text{orb}$. 
This can be accomplished by decomposing the regular representation into irreducibles, which in the case of abelian $G_r$ are all one-dimensional. In this way, we obtain a basis of $I^\text{orb}$, in which $G_r$ acts diagonally.
  
  Any finite abelian group is isomorphic to a product ${\mathbb Z}_{d_1}\times\ldots\times {\mathbb Z}_{d_r}$. 
  We will spell out the details for the case, in which it is isomorphic to a single factor 
$G_r\cong {\mathbb Z}_d$. The generalization to more factors is straight-forward.

A basis of $V_\text{reg}$ corresponding to the irreducible representations 
 can be obtained by performing the following transformation:
\begin{equation} \label{eq:diagonalbasis}
{e}_j = \sum_{g\in\bZ_d} \xi^{-gj} g
\,,\quad 0\leq j<d\,,
\end{equation}
where $\xi=\exp(2\pi i/d)$ is an elementary $d$th root of unity. $g\mapsto \xi^{jg}$ is the character of the irreducible representation $\rho_j$ defined by
 $\rho_j([n]_d)=\xi^{jn}$. Hence, 
$e_j$ is the basis vector of the irreducible representation $\rho_j([n]_d)=\xi^{jn}$, which is of course nothing but the $j$-fold tensor product of $\rho_1$: $\rho_j=\rho_1^{\otimes j}$. 
Thus, we can write $e_j=e_1^{\otimes j}$. Note that $\rho_1^{\otimes d}=\rho_0$. Writing $e_j=\alpha^{-j}$, 
the regular representation can be expressed as
\begin{equation}
V_\text{reg}\cong \bC[\alpha]/(\alpha^d-1)\,.
\end{equation}
Note that this is not only a vector space, but also a ring, and that multiplication in this ring
corresponds to taking the tensor product of representations. This allows us to rewrite the identity matrix factorization in the Landau-Ginzburg orbifold with orbifold group $G=\bZ_d$ as
\begin{equation}\label{module2}
I^\text{orb}=S_{(X)(X')}\otimes\Lambda(V)\otimes\bC[\alpha]/(\alpha^d-1)
\end{equation}
with differential
\[\begin{aligned}
	&\d_{I^\text{orb}} = \sum_{i=1}^n\left[ (X_i-\alpha^{Q_i} X_i')\cdot \theta^*_i + \partial^{X,\alpha X'}_{i}W\cdot \theta_i \right] \quad\text{with}\\
	&\partial^{X,\alpha X'}_iW = \frac{W(\alpha^{Q_1}X_1',..., \alpha^{Q_{i-1}}X_{i-1}', X_i, ..., X_n) - W(\alpha^{Q_1}X_1', ..., \alpha^{Q_i}X_{i}', X_{i+1}, ..., X_n)}{X_i-\alpha^{Q_i} X_{i}'}\,.
\end{aligned}\]
Here, $Q_i$ denote the charges of the chiral fields $X_j$ under the orbifold group $\bZ_d$, i.e.~$[n]\in\bZ_d$
acts on the chiral fields as $X_j\mapsto \xi^{Q_j n}X_j$. $\alpha$ can be regarded as a new bosonic defect field carrying charg $(1,-1)$ under the product $\bZ_d\times\bZ_d$ of the left and right orbifold groups.\footnote{The charge under the action of the right group is clear, because $\alpha$ represents the basis vector of the irreducible representation $\rho_{-1}$ under the right group. That it has charge $1$ under the left group follows because $V_\text{reg}$ was chosen to be invariant under the left-right diagonal subgroup.}

The representation on $I^\text{orb}$ under $G\times G$ is now completely fixed by the choice of a one-dimensional representation of the diagonal subgroup, since the latter left the identity defect of the non-orbifold theory invariant. We choose it to be trivial to obtain the identity defect in the orbifold theory. (Other choices lead to defects implementing the quantum symmetry of the orbifold theory.) The representations of $G\times G$ on the module (\ref{module2}) is determined by the representation on the chiral fields $X_i$, the $\theta_i$ (which transform like the $X_i$) and the representation on $\alpha^i$. 

Let us give an explicit example which will be important later. Consider the Landau-Ginzburg model with a single chiral superfield $X$, superpotential $W(X)=X^d$ and orbifold group $G=\bZ_d$. $[n]_d\in\bZ_d$ acts on $X$ by multiplication with a phase $X\mapsto e^{2\pi ind'\over d} X$ which leaves $W(X)$ invariant. ($X$ has charge $d'$ under $\bZ_d$.) Following the construction above, one obtains the identity matrix factorization
\begin{equation*}
{I}^\text{orb}:{S}^\alpha\{[1]_d,[0]_d\}\\ 
	\tikz[baseline=0]{
		\node at (0,.5) {$i_1=(X-\alpha^{d'} X')$};
		\draw[arrow position = 1] (-2,.2) -- (2,.2);
		\draw[arrow position = 1] (2,0) -- (-2,0);
		\node at (0,-.4) {$i_0=\prod_{i=1}^{d-1}(X-\xi^i\alpha^{d'} X')$};
	}\, {S}^\alpha\{[0]_d,[0]_d\}\,,
\end{equation*}
where ${S}^\alpha:=\bC[X,X',\alpha]/(\alpha^d-1)$, and the $\{\cdot,\cdot\}$ denote a shift in $\bZ_d\times\bZ_d$-charges.
The associated $C_{(X)(X')}$-module is given by
\begin{equation}\label{eq:lgidmodule}
{S}^\alpha/(X-\alpha^{d'} X') \ .
\end{equation}
One can unpack this, by replacing $\alpha$ by a cyclic shift matrix. This yields the equivalent representation
\begin{equation}\label{eq:lgidmf}
I^\text{orb}:
{S}^d
\left(\scriptsize
\begin{array}{c}
\{[1]_d,[0]_d\}\\
\{[2]_d,[-1]_d\}\\
\vdots\\
\{[d]_d,[-1+d]_d\}
\end{array}
\right)
\\ 
	\tikz[baseline=0]{
		\node at (0,.5) {$\imath_1=(X\bI_d - \epsilon^{d'} X')$};
		\draw[arrow position = 1] (-2,.2) -- (2,.2);
		\draw[arrow position = 1] (2,0) -- (-2,0);
		\node at (0,-.4) {$\imath_0=\prod_{i=1}^{d-1}(X\bI_d-\xi^i\epsilon^{d'} X')$};
	}\, 
	{S}^d
	\left(\scriptsize
\begin{array}{c}
\{[0]_d,[0]_d\}\\
\{[1]_d,[-1]_d\}\\
\vdots\\
\{[d-1]_d,[-1+d]_d\}
\end{array}
\right)
\end{equation}
Here $S=\bC[X,X']$, $\bI_d$ is the $d\times d$-identity matrix, and 
\begin{equation}\label{eq:shiftmatrix}
\epsilon_d=\left(
\begin{array}{cccc}
0&&&1\\
1&\ddots&&\\
&\ddots&\ddots&\\
&&1&0
\end{array}
\right)
\end{equation}
denotes the $d\times d$-shift matrix.

\subsubsection{The identity defect in abelian gauged linear sigma models} \label{sec:identity}

For gauged linear sigma models, mainly boundaries were considered in the literature, see \cite{Herbst:2008jq,Hori:2013ika,Knapp:2016rec}. Defects can in principle be discussed along the same lines, for example using the folding trick, see \cite{Brunner:2008fa,Honda:2013uca}. In \cite{Brunner:2008fa} identity defects of Landau-Ginzburg phases of GLSMs are lifted to GLSMs. Using such constructions, one cannot obtain an identity defect of the GLSM itself, because the lifts are matrix factorizations of finite rank.

We now use the method presented in the previous section to construct an identity defect for a $U(1)$-orbifold of a Landau-Ginzburg model with chiral fields $X_1,\ldots,X_n$ and superpotential $W\in\bC[X_1,\ldots,X_n]$. The action of $U(1)$ on the chiral fields is specified by their charges $(Q_1,\ldots,Q_n)$, where $\varphi\in U(1)$ acts on $X_j$ by $X_j\mapsto e^{2\pi i Q_j\varphi} X_j$. (Generalizations to higher rank abelian gauge groups are straight-forward.) 

Since the orbifold group is infinite (and not even countable), the method of images cannot be applied in this situation. As it turns out, the formulation with the additional defect field $\alpha$ however can be adapted. The irreducible representations of $U(1)$ are countable, $\rho_j(\varphi)=e^{2\pi i j\varphi}$, $j\in\bZ$, with $\rho_i\otimes\rho_j\cong \rho_{i+j}$. But in contrast to the case of representations of $\bZ_d$, not all representations can be obtained as tensor products of a single representation $\rho_{-1}$. One needs an additional representation $\rho_1$ to generate all representations by means of the tensor product. So in the $U(1)$-case, instead of one additional bosonic defect field $\alpha$, one has to introduce two fields $\alpha,\alpha^{-1}$  which are inverse to each other, i.e.~$\alpha\alpha^{-1}=1$. They carry $U(1)\times U(1)$-charges $(1,-1)$ and $(-1,1)$, respectively.

With these additional fields one can construct the following defect in complete analogy to (\ref{module2})
\begin{equation}\label{eq:continuous-id}
I=S_{(X),(X')}\otimes\Lambda(V)\otimes\bC[\alpha,\alpha^{-1}]/(\alpha\alpha^{-1}-1)
\end{equation}
with differential
\[\begin{aligned}
	&\d_I = \sum_{i=1}^n\left[ (X_i-\alpha^{Q_i} X_i')\cdot \theta^*_i + \partial^{X,\alpha X'}_{i}W\cdot \theta_i \right]\,, \quad\\
	&\partial^{X,\alpha X'}_iW = \frac{W(\alpha^{Q_1}X_1',..., \alpha^{Q_{i-1}}X_{i-1}', X_i, ..., X_n) - W(\alpha^{Q_1}X_1', ..., \alpha^{Q_i}X_{i}', X_{i+1}, ..., X_n)}{X_i-\alpha^{Q_i} X_{i}'}\,.
\end{aligned}\]
The $U(1)\times U(1)$-representation on this matrix factorization is completely determined by the transformation properties of the fields $X_i$, i.e.~their $U(1)$-charges $Q_i$. The $\theta_i$ transform as the $X_i$. 

Note that this matrix factorization is of infinite rank! 

As in the case of finite orbifolds, there is the possibility of shifting the charges of the orbifold group on one side, relative to the one on the other. The defect with such a shift implements a quantum symmetry. Since we are interested in the identity defect, we set this shift to zero.

As alluded to above, the generalization to orbifold groups $U(1)^k$ $k>1$ is straight-forward. One just has to introduce a pair of additional fields $(\alpha,\alpha^{-1})$ for each $U(1)$-factor. 

In the discussion of the identity defect of finite Landau-Ginzburg orbifolds in the last subsection, we just gave a 
different but equivalent representation of the known identity defect. In the case of the $U(1)$-orbifold, the defect (\ref{eq:continuous-id}) is new, and we have to show that it really is the identity defect, i.e. that under fusion it behaves as the  unit.

To do so, we use associated modules as explained in section~\ref{sec:lgid}. For this we have to introduce some  notation:
\begin{eqnarray}
S_{(X),(X')}&=&\bC[X_1,\ldots,X_n,X'_1,\ldots,X'_n]\nonumber\\
S_{(X),(X')}^{(\alpha,\alpha^{-1})}&=&S[\alpha,\alpha^{-1}]/(\alpha\alpha^{-1}-1)\nonumber\\
C_{(X),(X')}&=&S_{(X),(X')}/(W(X_1,\ldots,X_n)-W(X_1',\ldots,X_n'))\nonumber\\
 C_{(X),(X')}^{(\alpha,\alpha^{-1})}&=&C_{(X),(X')}[\alpha,\alpha^{-1}]/(\alpha\alpha^{-1}-1)\,.\nonumber
\end{eqnarray}
Replacing $X$ or $X'$ by $-$ in the subscripts means setting the respective variables to zero.

To the matrix factorization $I$ constructed above 
we associate the $C_{(X),(X')}$-module
\begin{equation}\label{eq:idmodule}
M_I= C_{(X),(X')}^{(\alpha,\alpha^{-1})}/\left(X_1- \alpha^{Q_1} X_1',\dots , X_n-\alpha^{Q_n} X_n'\right)\,.
\end{equation}
As in the discussion of the identity defects in unorbifolded Landau-Ginzburg models in section~\ref{sec:lgid},
this module has a free Koszul-type resolution, which after finitely many steps turns into the two-periodic complex induced by the matrix factorization $I$.

Let $P'$ be a $U(1)$-equivariant matrix factorization of $W(X_1',\ldots,X_n')$. The analysis of the fusion $I\otimes P'$ now runs in complete analogy of the discussion of the fusion of the identity defect in the unorbifolded Landau-Ginzburg models in section~\ref{sec:lgid}, except for the fact that the fusion in the orbifold corresponds only to the part of the tensor product matrix factorization which is invariant under the gauge group associated to the model squeezed in between the two defects. 

Let
$$
M_{P'}=\coker(p'_1:P'_1\otimes_{{S}_{(-)(X')}} C_{(-)(X')}\rightarrow P'_0\otimes_{S_{(-)(X')}} C_{(-)(X')})
$$
be the module associated to $P'$. The matrix factorization given by the fusion of $I$ and $P'$ can be extracted from the $U(1)$-invariant part $M^{U(1)}$ of the $C_{(X),(-)}$-module
$$
M=M_I\otimes_{C_{(-)(X')}}M_{P'}\,.
$$
The relations in (\ref{eq:idmodule}) can now be used to replace all $X_i'$ by $\alpha^{-Q_i}X_i$. This eliminates all the variables $X_i'$. Let us next choose generators  $e_r$ of $M_{P'}$, on which $U(1)$ acts diagonally, with respective $U(1)$-charge $q_r$. Then $M$ is generated by $\alpha^i\otimes e_r$, where $i\in\bZ$. Most of these generators are not $U(1)$-invariant though. Only the $\tilde{e}_r:=\alpha^{-q_r}\otimes e_r$ generate $M^{U(1)}$. Thus, for each generator $e_r$ of $M_{P'}$ of $U(1)$-charge $q_r$ there is exactly one generator of $M^{U(1)}$, which also has $U(1)$-charge $q_r$. (Recall that $\alpha$ has $U(1)\times U(1)$-charge $(1,-1)$.) The relations between the generators $e_r$ in $M_{P'}$ become relations between the respective generators $\tilde{e}_r$, where all the $X_i'$ are replaced by $X_i$. Thus, $M^{U(1)}$ is isomorphic to the module $M_P$ associated to the matrix factorization $P$ obtained from $P'$ by replacing all $X_i'$ by $X_i$. Therefore, fusion with $I$ maps matrix factorizations to equivalent ones, and the matrix factorization $I$ acts as the identity matrix factorization.

\section{\texorpdfstring{Example: GLSM with superpotential $W(X,P)=X^dP^{d'}$}{Example: GLSM with superpotential W(X,P)=XdPd'}}\label{sec:example}
\subsection{The model and its phases}
In this section  we  exemplify our method in a concrete example of a gauged linear sigma model with two Landau-Ginzburg phases. The model has a single  $U(1)$-gauge group and two chiral fields $X$ and $P$ 
of $U(1)$-charges $Q_x=d'$, respectively $Q_p=-d$. Its superpotential is given by $W=X^dP^{d'}$. 
We assume $d>d'$, and for simplicity we restrict to the case 
where $d$ and $d'$ are coprime integers.

For this model, the scalar potential \eqref{eq:general scaler potential} takes the form
\[
	U = |\sigma|^2 \left( Q_x^2 |x|^2 + Q_p^2 |p|^2\right) + \frac{e^2}2 \left( Q_x |x|^2 + Q_p |p|^2 - r  \right)^2 + \left\vert \partial_x W(p,x)\right\vert^2 + \left\vert \partial_p W(p,x)\right\vert^2.
\]
The total charge $Q_{tot}=d'-d<0$ is negative, which means that the Fayet-Iliopoulos parameter (\ref{eq:running}) runs under the RG flow, from $r<<0$ in the UV to $r>>0$ in the IR. The model exhibits two Landau-Ginzburg phases. 

For $r<0$, the D-term constraint coming from the second term above, requires $p\neq 0$. This breaks the $U(1)$ gauge symmetry to $\bZ_{|Q_p|}=\bZ_d$ and $\sigma$ must vanish according to the first term. Because of the first superpotential term, $x$ also vanishes and hence $|p|^2=\frac{r}{Q_p}=-\frac{r}{d}$. We obtain a Landau-Ginzburg orbifold model with one chiral field $X$, superpotential $X^d$ and orbifold group $\bZ_d$. This model is well known to give a Landau-Ginzburg realization of the $N=(2,2)$ minimal superconformal model with central charge $c=(d-2)/d$. 

For $r>0$, the roles of $X$ and $P$ are interchanged. The D-term constraint yields $x\neq 0$, which further implies that $\sigma$ and $p$ vanish. The $U(1)$ gauge group is broken to $\bZ_{|Q_x|}=\bZ_{d'}$, and $|x|^2 = \frac{r}{Q_x}=\frac{r}{d'}$. We arrive at a Landau-Ginzburg orbifold model with chiral field $P$, superpotential $P^{d'}$ and orbifold group $\bZ_{d'}$. Again, this yields a Landau-Ginzburg realization fo the $N=(2,2)$ minimal model with the smaller central charge $c=(d'-2)/d'$, to which the systems flows at low energies.

Classically, there is a Coulomb branch emerging at $r=0$, parametrized by $\sigma$. Due to a twisted superpotential, the values of $\sigma$ will be restricted to a finite set of  $d-d'$ massive vacua that appear in the IR phase.

In the following we will use our general strategy to construct defects describing the transitions between UV and IR phase of this model, defects embedding the two phases in the GLSM as well as defects projecting the GLSM to the phases.

Note that there is an effective description of the mirror of this GLSM in terms of an ordinary Landau-Ginzburg model \cite{Vafa:2001ra}, namely the Landau-Ginzburg model with one chiral field $X$ and superpotential
$$
W=X^d+e^{t\over d} X^{d'}\,.
$$
The deformation parameter $\lambda=e^{t\over d}$ of the superpotential is related to the complexified Fayet-Iliopoulus parameter $t=r-i\theta$ of the GLSM. $\lambda$ runs under the RG flow from $\lambda=0$ in the UV to $\lambda=\infty$ in the IR. In the UV, the model is therefore described by a LG model with superpotential $W=X^d$ and in the IR by a LG model with superpotential $W=X^{d'}$. 

Now, flows of Landau-Ginzburg models triggered by deformations of the superpotentials are relatively well under control, and at least some aspects of them can be studied very explicitly. For instance, it is not difficult to analyze what happens to the vacua of the model, which correspond to critical points of the superpotential. In the case at hand,  some of these vacua ($d-d'$ many) move off to infinity under the RG flow, and decouple from the theory, taking with them some (A-type) D-branes attached to them. 

This decoupling of D-branes is well described by RG defects associated to the flows. Indeed, all the RG defects corresponding to flows of Landau-Ginzburg models with a single chiral superfield but general deformations of the superpotential
\begin{equation}\label{eq:perturbed}
W=X^d +\sum_{i=1}^{d-1} \lambda_i X^i \ ,
\end{equation}
have been constructed in \cite{Brunner:2007ur}.\footnote{More precisely, they have been constructed in the mirror theories.}
Thus, the transition defects between UV and IR phases of the GLSM which we will obtain here can be checked against known results. We find complete agreement.

\subsection{GLSM identity defect}

The starting point of our analysis is the identity defect of the GLSM as constructed for the general abelian GLSMs in section~\ref{sec:identity}. In this case, it is a $U(1)\times U(1)$-equivariant matrix factorization of the difference $W(X,P)-W(Y,Q)=X^dP^{d'}-Y^dQ^{d'}$ of the superpotentials of the gauged linear sigma models on either side of the defects.
The two $U(1)$-factors correspond to the gauge groups of the models on the left and the right of the defect, respectively.

For clarity, we will repeat and spell out some details of the construction of the identity defect in this example. 
We first introduce new variables (corresponding to degrees of freedom on the identity defect) $\alpha$ and $\alpha^{-1}$ which satisfy $\alpha\alpha^{-1}=1$ and which carry $U(1)\times U(1)$-charges 
$|\alpha|=(1,-1)$ and $|\alpha^{-1}|=(-1,1)$. 
Using these fields, we can write the difference of the superpotentials as follows
\begin{eqnarray}
W(X,P)-W(Y,Q)&=& X^dP^{d'}-Y^dQ^{d'}=X^dP^{d'}-\alpha^{dd'}Y^dP^{d'}+\alpha^{dd'}Y^dP^{d'}-Y^dQ^{d'}\nonumber\\
&=&\left(X^d-(\alpha^{d'}Y)^d\right)\,P^{d'}+Y^d\,\left((\alpha^dP)^{d'}-Q^{d'}\right)\nonumber\\
&=&P^{d'}\,\prod_{i=0}^{d-1}(X-\xi^i\alpha^{d'}Y) + Y^d\prod_{i=0}^{d'-1}(\alpha^dP-(\xi')^iQ)\,.\nonumber
\end{eqnarray}
Here $\xi=e^{2\pi i \over d}$ and ${\xi}'=e^{2\pi i \over d'}$ are elementary $d$th, respectively $d'$th roots of unity. 

The matrix factorization associated to the identity defect is then given by the Koszul-type matrix factorization associated to 
$(X-\alpha^{d'}Y)$ and $(\alpha^dP-Q)$. More precisely, denoting the $\bC[X,P,Y,Q]$-modules
\begin{eqnarray}
S&=&S_{(X,P)(Y,Q)}=\bC[X,P,Y,Q]\nonumber\\
\text{and}\;\;\;\tilde{S}&=&{S}^{(\alpha,\alpha^{-1})}_{(X,P)(Y,Q)}=S_{(X,P)(Y,Q)}[\alpha,\alpha^{-1}]/(\alpha\alpha^{-1}-1)\,.\nonumber
\end{eqnarray}
the identity matrix factorization can be written as 
\begin{equation}\label{eq:idmf}
I:\tilde{S}^2\left(\begin{array}{c}\{d',0\}\\\{0,-d\}\end{array}\right)\\ 
	\tikz[baseline=0]{
		\node at (0,.5) {$i_1$};
		\draw[arrow position = 1] (-1.5,.2) -- (1.5,.2);
		\draw[arrow position = 1] (1.5,0) -- (-1.5,0);
		\node at (0,-.4) {$i_0$};
	}\, \tilde{S}^2\left(\begin{array}{c}\{0,0\}\\\{d',-d\}\end{array}\right)
\end{equation}
Here $\{\cdot,\cdot\}$ indicates the $U(1)\times U(1)$-charge of the respective generator and 
\begin{eqnarray}
i_1&=&\left(
\begin{array}{cc}
(X-\alpha^{d'}Y) & -(\alpha^dP-Q)\\
Y^d\prod_{i=1}^{d'-1}(\alpha^dP-(\xi')^iQ) &
P^{d'}\prod_{i=1}^{d-1}(X-\xi^i\alpha^{d'}Y)
\end{array}
\right)
\nonumber\\
i_0&=&\left(
\begin{array}{cc}
 P^{d'}\prod_{i=1}^{d-1}(X-\xi^i\alpha^{d'}Y)& (\alpha^dP-Q)\\
-Y^d\prod_{i=1}^{d'-1}(\alpha^dP-(\xi')^iQ) &
(X-\alpha^{d'}Y) \  .
\end{array}
\right)\nonumber
\end{eqnarray}
This is nothing but the GLSM identity matrix factorization (\ref{eq:continuous-id}) spelled out for the special case at hand.
To it we associate the module
\begin{equation}\label{eq:idmod}
M_{I}=C_{(X,P)(Y,Q)}^{(\alpha,\alpha^{-1})}/((X-\alpha^{d'}Y),(\alpha^dP-Q))
\end{equation}
over the ring
\begin{equation}
C=C_{(X,P)(Y,Q)}=S_{(X,P)(Y,Q)}/(W(X,P)-W(Y,Q))\,,\nonumber
\end{equation}
where
\begin{equation}
{C}^{(\alpha,\alpha^{-1})}_{(X,P)(Y,Q)}=C_{(X,P)(Y,Q)}[\alpha,\alpha^{-1}]/(\alpha\alpha^{-1}-1)\,,\nonumber
\end{equation}
c.f.~the general case (\ref{eq:idmodule}). 
The Koszul resolution of $M_I$ turns into  the two-periodic complex induced by the identity matrix factorization $I$ after two steps.

\subsection{Pushing down the identity defect into phases}\label{sec:glsmidtolgid}
Going into the phases of the GLSM, one of the two chiral fields gets a vacuum expecation value (which we can take to be $1$), and the gauge group is broken to the subgroup leaving this chiral field invariant. We can therefore push down any defect of the GLSM into a phase by setting the respective chiral field to $1$ in the associated matrix factorization and considering it equivariant with respect to the residual gauge group. In fact, this can be done on either side of the defect yielding defects from the GLSM into the phases or from the phases to the GLSM.
Moreover, one can push down to phases on both sides of the defect, possibly into different phases on the two sides, which gives rise to defects in the phases or from one phase to another.

In the next section, we will employ this push-down to the GLSM identity defect. Pushing down to the UV phase on the right side and the IR phase on the left,  we will obtain a defect describing the transition from the UV phase to the IR phase. Indeed, we will reproduce RG defects of \cite{Brunner:2007ur}, as expected.

Before we will come to this, as a warm-up we first discuss the simpler case, where the GLSM identity defect is pushed down to the same phase on both sides. We choose the UV phase. The push down to the IR can be dealt with in a similar way.

To push down the GLSM identity defect  to the UV phase on both sides, we have to set $P$ and $Q$ to $1$ in the matrix factorization (\ref{eq:idmf}) and consider it equivariant with respect to the residual gauge group
$\bZ_d\times\bZ_d$. We can do this on the level of the associated module (\ref{eq:idmod}). 

It will be useful to introduce some notation. Replacing the name of a variable with a `$\cdot$' in the subscripts of the rings
$S_{(X,P)(Y,Q)}$, $C_{(X,P)(Y,Q)}$ or ${C}^{(\alpha,\alpha^{-1})}_{(X,P)(Y,Q)}$ just means setting the respective variable to one.\footnote{Or to put it differently, by diving the rings by the corresponding ideals.} For instance
\begin{eqnarray}
S_{(X,\cdot)(Y,Q)}&=&\bC[X,Y,Q]\nonumber\\
C_{(X,\cdot)(Y,Q)}&=&S_{(X,\cdot)(Y,Q)}/(W(X,1)-W(Y,Q))\nonumber\\
{C}^{(\alpha,\alpha^{-1})}_{(X,\cdot)(Y,Q)}&=&C_{(X,\cdot)(Y,Q)}[\alpha,\alpha^{-1}]/(\alpha\alpha^{-1}-1)\,.\nonumber
\end{eqnarray}
Pushing down the GLSM identity defect to the UV phase on both sides yields the module
\begin{equation*}
M_I^{\rm UV\, UV}={C}^{(\alpha,\alpha^{-1})}_{(X,\cdot)(Y,\cdot)}/((X-\alpha^{d'}Y),(\alpha^d-1))
\,.
\end{equation*}
Note that due to the relation $\alpha^d-1$ this module is of finite rank, and in fact isomorphic to the module (\ref{eq:lgidmodule}) associated to the identity matrix factorization of the LG model describing the UV phase. In fact, 
identifying\footnote{the generator $\alpha^i$ is sent to the generator $e_i$}
\begin{equation*}
{C}^{(\alpha,\alpha^{-1})}_{(X,\cdot)(Y,\cdot)}/(\alpha^d-1)
\cong
{C^d_{(X,\cdot)(Y,\cdot)}}\left(
\begin{array}{c}
\{[0]_d,[0]_d\}\\
\{[1]_d,[-1]_d\}\\
\vdots\\
\{[d-1]_d,[-d+1]_d\}
\end{array}
\right)\,,
\end{equation*} 
where $\{\cdot,\cdot\}$ denotes the shift in $\bZ_d\times\bZ_d$-charge of the respective generators,
one can write $M_{I}^{\rm UV\,UV}$  as cokernel 
\begin{equation*}
M_{I}^{\rm UV\,UV}\cong
\coker\left(\imath^{\rm UV}_1:
{C}^d_{(X,\cdot)(Y,\cdot)}\left({\tiny
\begin{array}{c}\scriptstyle
\{[d']_d,[0]_d\}\\
\{[d'+1]_d,[-1]_d\}\\
\vdots\\
\{[d'+d-1]_d,[-d+1]_d\}
\end{array}}
\right)
\rightarrow{C}^d_{(X,\cdot)(Y,\cdot)}\left({\tiny
\begin{array}{c}
\{[0]_d,[0]_d\}\\
\{[1]_d,[-1]_d\}\\
\vdots\\
\{[d-1]_d,[-d+1]_d\}
\end{array}}
\right)\right)
\end{equation*}
of the map $\imath_1^{\rm UV}=(X\bI_d-Y\epsilon_d^{d'})$. 
Here, as before $\bI_d$ denotes the $d\times d$-identity matrix and $\epsilon_d$ the $d\times d$-shift matrix (\ref{eq:shiftmatrix}).

Indeed, $\imath_1^{\rm UV}\imath_0^{\rm UV}=(X^d-Y^d)\bI_d$ for 
\begin{equation*}
\imath_0^{\rm UV}=\prod_{i=1}^{d-1}(X\bI_d-\xi^iY\epsilon_d^{d'})\,.
\end{equation*}
Hence, $\imath_1^{\rm UV}$ is a factor of a matrix factorization of $W(X,1)-W(Y,1)$, namely
\begin{equation}\label{eq:uvid}
I^{\rm UV}:
{S}^d_{(X,\cdot)(Y,\cdot)}\left(\tiny
\begin{array}{c}
\{[1]_d,[0]_d\}\\
\{[2]_d,[-1]_d\}\\
\vdots\\
\{[d]_d,[-d+1]_d\}
\end{array}
\right)
\\ 
	\tikz[baseline=0]{
		\node at (0,.5) {$\imath^{\rm UV}_1$};
		\draw[arrow position = 1] (-1.5,.2) -- (1.5,.2);
		\draw[arrow position = 1] (1.5,0) -- (-1.5,0);
		\node at (0,-.4) {$\imath^{\rm UV}_0$};
	}\, 
	{S}^d_{(X,\cdot)(Y,\cdot)}\left(\tiny
\begin{array}{c}
\{[0]_d,[0]_d\}\\
\{[1]_d,[-1]_d\}\\
\vdots\\
\{[d-1]_d,[-d+1]_d\}
\end{array}
\right)
\end{equation}
This matrix factorization corresponds to the identity defect (\ref{eq:lgidmf}) in the UV phase.
Thus, the module $M_I^\text{UV UV}$ is associated to the identity matrix factorization in the UV phase.
Pushing down the GLSM identity defect on both sides to the UV, one therefore produces the identity defect in the UV phase. 
Similarly, pushing down the GLSM identity defect to the IR phase on both sides yields the IR identity defect.
This is of course what is to be expected.

\subsection{RG defects from the GLSM identity}
Next, we push down the GLSM identity defect to the UV on the right and to the IR on the left, to obtain a transition defect between UV and IR phase. Note that there is more than one (homotopy class of) paths from UV to IR. So there should be more than one such transition defects. 

Implementing the push-down involves setting those variables to $1$ in the GLSM identity matrix factorization, which correspond to fields acquiring a vacuum expectation value in the respective phases. These are $X$ (IR phase on the left of defect) and $Q$ (UV phase on the right).
On the level of $C_{(\cdot,P)(Y,\cdot)}$-modules this yields
\begin{equation}\nonumber
M_I^{\rm UV\,IR}=
C^{(\alpha,\alpha^{-1})}_{(\cdot,P)(Y,\cdot)}/((Y-\alpha^{-d'}),(P-\alpha^{-d}))\,.
\end{equation}
In contrast to the push down to the same phase on both sides, this module is of infinite rank, leading to a matrix factorization of infinite rank, which does not correspond to one of the RG defects between the respective Landau-Ginzburg models. We propose that under the push-down to the phases, 
the module (and the respective matrix factorization) has to be truncated to finite rank. The truncation is not unique, but it turns out, that the different choices of truncation exactly correspond to the different paths from UV to IR.\footnote{We expect the truncation to be related to the gradability of the resulting matrix factorization with respect to R-symmetry. The latter ensures definite gluing conditions for the spectral flow operators of the respective SCFTs along the defect. This is needed to impose a stability condition in the sense of \cite{Walcher:2004tx} on the level of the defect.}
Concretely, we introduce an upper bound $N$ on the $\alpha$-exponents\footnote{Later we will show that $N$ determines the charge window for the grade restriction rule which appears in \cite{Herbst:2008jq}.}
\begin{equation}\nonumber
M_I^{\rm UV\,IR}(N)=
\alpha^{N}C_{(\cdot,P)(Y,\cdot)}[\alpha^{-1}]/((Y-\alpha^{-d'}),(P-\alpha^{-d}))\alpha^N{C}_{(\cdot,P)(Y,\cdot)}[\alpha^{-1}].
\end{equation}
This module is now of finite rank.
It has generators $e_i:=\alpha^{N-i}$ for $0\leq i<d'$ of $\bZ_{d'}\times\bZ_{d}$-charges $([N-i]_{d'},-[N-i]_d)$. To write down the relations in a convenient way we define $a,b\in\bN$ with $b<d'$ such that
\begin{equation}
d=ad'+b\,.
\end{equation}
Then the generators satisfy relations
\begin{eqnarray}
Pe_i&=&P\alpha^{N-i}=\alpha^{N-d-i}=Y^a\alpha^{N-d-i+ad'}\nonumber\\
&=&\left\{\begin{array}{ll}Y^ae_{i+b}\,,&i+b<d'\\Y^{a+1}e_{i+b-d'}\,,&i+b\geq d'\end{array}\right.\nonumber
\end{eqnarray}
and the module $M_{I}^{\rm UV\,IR}(N)$ is isomorphic to the cokernel of a map
\begin{equation*}
rg_1:C^{d'}_{(\cdot,P)(Y,\cdot)}
\left({\tiny
\begin{array}{c}
\{[N-d]_{d'},[-N]_d\}\\
\{[N-d-1]_{d'},[-N+1]_d\}\\
\vdots\\
\{[N-d-d'+1]_{d'},[-N-1+d']_d\}
\end{array}}
\right)
\rightarrow
C^{d'}_{(\cdot,P)(Y,\cdot)}
\left({\tiny
\begin{array}{c}
\{[N]_{d'},[-N]_d\}\\
\{[N-1]_{d'},[-N+1]_d\}\\
\vdots\\
\{[N-d'+1]_{d'},[-N+d'-1]_d\}
\end{array}}
\right)\,,
\end{equation*}
which can be written as $rg_1=(P\bI_{d'}-\epsilon_{d'}^bI_Y)$, where $I_Y$ denotes the 
diagonal $d'\times d'$-matrix with $Y^a$ as its first $d'-b$ diagonal entries and $Y^{a+1}$ as the last $b$ diagonal entries. Explicitly
\begin{equation*}
rg_1=\left(\begin{array}{cccccc}
P&&&-Y^{a+1}&&\\
&\ddots&&&\ddots&\\
&&\ddots&&&-Y^{a+1}\\
-Y^a&&&\ddots&&\\
&\ddots&&&\ddots&\\
&&-Y^a&&&P
\end{array}\right)\,.
\end{equation*}
Now 
\begin{equation*}
\prod_{i=0}^{d'-1}(P\bI_{d'}-(\xi')^i\epsilon_{d'}^b I_Y)=(P^{d'}-Y^d)\bI_{d'}\,,
\end{equation*}
and hence, $rg_1$ together with $rg_0=\prod_{i=1}^{d'-1}(P\bI_{d'}-(\xi')^i\epsilon_{d'}^b I_Y)$
defines a matrix factorization $RG_N$:
\begin{equation}\label{eq:grades}
{S}^{d'}_{(\cdot,P)(Y,\cdot)}\left({\tiny
\begin{array}{c}
\{[N-d]_{d'},-[N]_d\}\\
\{[N-1-d]_{d'},-[N-1]_d\}\\
\vdots\\
\{[N-d'+1-d]_{d'},-[N-d'+1]_d\}
\end{array}}
\right)
\\ 
	\tikz[baseline=0]{
		\node at (0,.5) {$rg_1$};
		\draw[arrow position = 1] (-0.3,.2) -- (0.3,.2);
		\draw[arrow position = 1] (0.3,0) -- (-0.3,0);
		\node at (0,-.4) {$rg_0$};
	}\, 
	{S}^{d'}_{(\cdot,P)(Y,\cdot)}\left({\tiny
\begin{array}{c}
\{[N]_{d'},-[N]_d\}\\
\{[N-1]_{d'},-[N-1]_d\}\\
\vdots\\
\{[N-d'+1]_{d'},-[N-d'+1]_d\}
\end{array}}
\right)
\end{equation}
Pushing down the GLSM identity defect to the UV on the right and to the IR on the left with truncation $N$ yields 
a defect between the Landau-Ginzburg models in the UV and in the IR,  given by 
the matrix factorization $RG_N$. Note that $N$ only appears in the grading of the matrix factorization $RG_N$, and that $RG_N=RG_{N+dd'}$. Furthermore, the shift in $N$ corresponds to conjugation with the quantum symmetries of the respective Landau-Ginzburg phases, $RG_{N+1}=Q_\text{IR}^{-1}\otimes R_N\otimes Q_\text{UV}$.

Thus, we obtain $dd'$ many different transitions defects between the two phases. 
These indeed correspond to particular 
renormalization group defects between Landau-Ginzburg orbifolds describing the  UV and IR phases \cite{Brunner:2007ur}.  
In fact  RG defects between these Landau-Ginzburg orbifolds corresponding to general perturbations of type (\ref{eq:perturbed}) would allow for a more generic distribution of powers of $Y$ in the map $rg_1$, of the form
\begin{equation*}
rg^{gen}_1=\left(\begin{array}{cccccc}
P&&&-Y^{n_b}&&\\
&\ddots&&&\ddots&\\
&&\ddots&&&-Y^{n_{d'}}\\
-Y^{n_1}&&&\ddots&&\\
&\ddots&&&\ddots&\\
&&-Y^{n_{b-1}} &&&P
\end{array}\right)\,.
\end{equation*}
where $\sum n_a=d$ and the grades appearing in (\ref{eq:grades}) have to be modified accordingly, c.f.~\cite{Brunner:2007ur}. The transition defects we obtain  from the gauged linear sigma model are special cases of these defects which exhibit a maximally homogeneous distribution of powers of $Y$ in $rg_1$.

Summarizing, by pushing down to UV and IR on the right, respectively left side of the GLSM identity defect with an additional truncation we obtain an RG defect between the UV and IR phases, which is known to describe the transport between UV and IR Landau-Ginzburg models. 
Different choices of the truncation parameter $N$ only shift the charges of the matrix factorization, in particular
$RG_N\cong RG_0\{[N]_{d'},-[N]_d\}\cong Q_\text{IR}^N\otimes RG_0\otimes Q_\text{UV}^{-N}$. The charge shifts are quantum symmetries of the Landau-Ginzburg orbifolds  in IR and UV. They can be obtained as monodromies of the GLSM upon encircling the Landau-Ginzburg points in the K\"ahler parameter space. Thus, up to winding around the limit points, we obtain one defect describing the flow between UV and IR phases of the GLSM.

\subsection{Factorization of RG defects}

Using the fact that the GLSM identity defect is an idempotent,
\begin{equation}\label{eq:ididempotent}
I\cong I\otimes I\,,
\end{equation}
we can factorize the RG defects $RG_N$ over the GLSM. More precisely
\begin{equation*}
{RG}_N\cong R^{\rm IR}\otimes T_N^{\rm UV}\,,
\end{equation*}
where $R^{\rm IR}$ is a defect from the GLSM to the IR phase obtained by pushing down the GLSM identity defect to the IR on the left side, and 
$T^{\rm UV}_N$ is a defect from the UV phase to the GLSM model obtained by pushing the GLSM identity defect to the UV on the right and truncating.\footnote{Note that the truncation could also be implemented on $R$ instead, or on both $R$ and $T$. As it turns out, pushing the truncation on $T$ and not on $R$ leads to a nice interpretation of the factor defects.}
Let us discuss the factor defects in turn.

\subsubsection{\texorpdfstring{$T_N^{\rm UV}$}{TNUV}}

The module associated to $T_N^{\rm UV}$ is obtained by setting $Q=1$ in (\ref{eq:idmod}) and then truncating the $\alpha$-spectrum. This yields the $C_{(X,P)(Y,\cdot)}$-module
\begin{equation}\nonumber
M_{I}^{\rm UV\,GLSM}(N)=
\alpha^{N}C_{(X,P)(Y,\cdot)}[\alpha^{-1}]/((Y-\alpha^{-d'}X),(P-\alpha^{-d}))\alpha^N{C}_{(X,P)(Y,\cdot)}[\alpha^{-1}].
\end{equation}
It is finitely generated with generators $e_i=\alpha^{N-i}$, of $U(1)\times\bZ_d$-charge $(N-i,-[N-i]_d)$, where $0\leq i<d$. The generators satisfy relations
\begin{eqnarray}\label{eq:TNrelations}
Ye_i&=&Y\alpha^{N-i}=X\alpha^{N-i-d'}\nonumber\\
&=&\left\{\begin{array}{ll}
Xe_{i+d'}\,,&i+d'<d\\
PXe_{i+d'-d}\,,&i+d'\geq d
\end{array}\right.\,,
\end{eqnarray}
and 
$M_{I}^{\rm UV\,GLSM}(N)$ is isomorphic to the cokernel of the map
\begin{equation*}
t_1:
C^{d}_{(X,P)(Y,\cdot)}
\left(
{\footnotesize
\begin{array}{c}
\{N,-[N-d']_d\}\\
\{N-1,-[N-1-d']_d\}\\
\vdots\\
\{N-d+1,-[N-d+1-d']_d\}
\end{array}}
\right)
\rightarrow
C^{d}_{(X,P)(Y,\cdot)}
\left({\footnotesize
\begin{array}{c}
\{N,[-N]_d\}\\
\{N-1,-[N-1]_d\}\\
\vdots\\
\{N-d+1,-[N-d+1]_d\}
\end{array}}
\right)\,,
\end{equation*}
with $t_1=(\epsilon_d^{d'}I_P X- Y\bI_d)$. Here $I_P$ is the diagonal $d\times d$-matrix with $1$ in the first $b=d-d'$ diagonal entries and $P$ in the last $d'$ diagonal entries. Explicitly,
\begin{equation*}
t_1=\left(\begin{array}{cccccc}
-Y&&&PX&&\\
&\ddots&&&\ddots&\\
&&\ddots&&&PX\\
X&&&\ddots&&\\
&\ddots&&&\ddots&\\
&&X&&&-Y
\end{array}\right)\,.
\end{equation*}
Now,
\begin{equation*}
\prod_{i=0}^{d-1}(\epsilon_d^{d'}I_P X- \xi^iY\bI_d)=P^{d'}X^d-Y^d\,,
\end{equation*}
and hence, together with the map
$t_0=\prod_{i=1}^{d-1}(\epsilon_d^{d'}I_P X- \xi^iY\bI_d)$, $t_1$ defines a matrix factorization of $W(X,P)-W(Y,1)$. 
Thus, the module $M_{I}^{\rm UV\,GLSM}(N)$ has a free two-periodic resolution induced by the matrix factorization
\begin{equation}\label{eq:tnuv}
T^{\rm UV}_N:
{S}^{d}_{(X,P)(Y,\cdot)}\left({\tiny
\begin{array}{c}
\{N,-[N-d']_d\}\\
\{N-1,-[N-1-d']_d\}\\
\vdots\\
\{N-d+1,-[N-d+1-d']_d\}
\end{array}}
\right)
\\ 
	\tikz[baseline=0]{
		\node at (0,.5) {$t_1$};
		\draw[arrow position = 1] (-0.3,.2) -- (0.3,.2);
		\draw[arrow position = 1] (0.3,0) -- (-0.3,0);
		\node at (0,-.4) {$t_0$};
	}\, 
	{S}^{d}_{(X,P)(Y,\cdot)}\left({\tiny
\begin{array}{c}
\{N,-[N]_d\}\\
\{N-1,-[N-1]_d\}\\
\vdots\\
\{N-d+1,-[N-d+1]_d\}
\end{array}}
\right)\,.
\end{equation}
This matrix factorization represents the  factor defect $T_N^{\rm UV}$. 
Note that it is of rank $d$ and exhibits $U(1)$-charge shifts only in the set $\{N-d+1,N-d+2,\ldots, N\}$ of $d$ consecutive integers starting at $N-d+1$. Fusion with the defect 
$T_N^{\rm UV}$ therefore lifts D-branes from the UV phase to GLSM branes in the charge window $\{N-d+1,\ldots,N\}$ in the terminology of \cite{Herbst:2008jq}.

Indeed, there is another way to arrive at the defects $T_N^{\rm UV}$. One can start with the identity defect in the UV phase and then lift on the left to the GLSM. Lifting in this case means inserting variables $P$ into the rank-$d$ $\bZ_d\times\bZ_d$-equivariant matrix factorization of $X^d-Y^d$ in such a way as to make it into a $U(1)\times\bZ_d$-equivariant matrix factorization of $P^{d'}X^d-Y^d$. (Lifting D-branes in such a manner is an important ingredient in the discussion of D-brane transfer between LG and geometric phases of abelian GLSMs in \cite{Herbst:2008jq}.) In this way one obtains a defect from the UV phase to the GLSM. This defect is automatically of finite rank, so a truncation of the kind we had to impose when coming from the GLSM is not necessary. On the other hand, the lift involves many choices. One of the choices corresponds to the choice of $N$, the maximal $U(1)$-charge. When that is fixed there are still choices left, and only one of them leads to the defects $T_N^{\rm UV}$. In fact, $T_N^{\rm UV}$ corresponds to the unique lift of the UV identity defect, which has maximal $U(1)$-charge $N$, and whose $U(1)$-charges populate $\{N-d+1,\ldots,N\}$. That means it is the only such lift, which upon fusion sends all UV branes to GLSM branes in the respective charge window of length 
$d$ in the terminology of \cite{Herbst:2008jq}.

As a side remark, pushing down the defect $T^{\rm UV}_N$ on the left to the IR (setting $X=1$), one obtains the RG defect $RG_N$. Pushing down on the left to the UV (setting $P=1$), yields the identity defect in the UV phase.

Of course, defects $T_N$ can be constructed for any phase. Pushing the GLSM identity defect to the IR on the right, in a similar fashion yields defects 
$T^{\rm IR}_N$ from the IR phase to the GLSM.

\subsubsection{\texorpdfstring{$R^{\rm IR}$}{RIR}}

$R^{\rm IR}$ is obtained by  pushing down to the IR on the left of the GLSM identity defect. Pushing down on the level of modules yields the ${C}_{(\cdot,P)(Y,Q)}$-module
\begin{equation*}
M_{I}^{\rm GLSM\,IR}={C}_{(\cdot,P)(Y,Q)}^{(\alpha,\alpha^{-1})}/
((Y-\alpha^{-d'}),(P -\alpha^{-d}Q))\,.
\end{equation*}
This module is of infinite rank. Note that $Y$ is invertible in this module. One way to look at it is as a limit of truncated modules
\begin{equation*}
M_{I}^{\rm GLSM\,IR}=\lim_{N\to\infty}M_{I}^{\rm GLSM\,IR}(N)
\end{equation*}
with
\begin{equation*}
M_{I}^{\rm GLSM\,IR}(N)=
\alpha^N{C}_{(\cdot,P)(Y,Q)}[\alpha^{-1}]/
((Y-\alpha^{-d'}),(P -\alpha^{-d}Q))\alpha^N{C}_{(\cdot,P)(Y,Q)}[\alpha^{-1}]\,.
\end{equation*}
The truncated module is finitely generated with generators $e_i=\alpha^{N-i}$, $0\leq i<d'$ of $\bZ_{d'}\times U(1)$-charges $([N-i]_{d'},-N+i)$. 
They satisfy relations
\begin{eqnarray}
Pe_i&=&P\alpha^{N-i}=Q\alpha^{N-i-d}=Y^aQ\alpha^{N-i-b}\nonumber\\
&=&\left\{\begin{array}{ll} Y^aQ e_{i+b}\,,&i+b<d'\\
Y^{a+1}Qe_{i+b-d'}\,,& i+b\geq d'\end{array}\right.\nonumber
\end{eqnarray}
Therefore, $M_{I}^{\rm GLSM\,IR}(N)$ is isomorphic to the cokernel of the map
\begin{equation*}
r_1^{\rm IR}:C^{d'}_{(\cdot,P)(Y,Q)}
\left({\tiny
\begin{array}{c}
\{[N-d]_{d'},-N\}\\
\{[N-1-d]_{d'},-N+1\}\\
\vdots\\
\{[N-d'+1-d]_{d'},-N+d'-1\}
\end{array}}
\right)
\rightarrow
C^{d'}_{(\cdot,P)(Y,Q)}
\left({\tiny
\begin{array}{c}
\{[N]_{d'},-N\}\\
\{[N-1]_{d'},-N+1\}\\
\vdots\\
\{[N-d'+1]_{d'},-N+d'-1\}
\end{array}}
\right)\,,
\end{equation*}
with $r_1^{\rm IR}=(P\bI_{d'}-Q\epsilon_{d'}^bI_Y)$. Here $I_Y$ is the $d'\times d'$-diagonal matrix with $Y^a$ as the first $d'-b$ diagonal entries and $Y^{a+1}$ as the last $b$ diagonal entries. 
Explicitly,
\begin{equation*}
r_1^{\rm IR}=\left(\begin{array}{cccccc}
P&&&-QY^{a+1}&&\\
&\ddots&&&\ddots&\\
&&\ddots&&&-QY^{a+1}\\
-QY^a&&&\ddots&&\\
&\ddots&&&\ddots&\\
&&-QY^a&&&P
\end{array}\right)\,.
\end{equation*}
Now,
$\prod_{i=0}^{d'-1}(P\bI_{d'}-(\xi')^iQ\epsilon_{d'}^bI_Y)=P^{d'}-Y^dQ^{d'}=:r_1^{\rm IR}\,r_0^{\rm IR}$, and therefore the truncated modules have two-periodic resolutions induced by the matrix factorizations
\begin{equation*}
R^{\rm IR}_N:
{S}^{d'}_{(X,P)(Y,\cdot)}\left({\tiny
\begin{array}{c}
\{[N-d]_{d'},-N\}\\
\{[N-1-d]_{d'},-N+1\}\\
\vdots\\
\{[N-d'+1-d]_{d'},-N+d'-1\}
\end{array}}
\right)
\\ 
	\tikz[baseline=0]{
		\node at (0,.5) {$r_1^{\rm IR}$};
		\draw[arrow position = 1] (-0.3,.2) -- (0.3,.2);
		\draw[arrow position = 1] (0.3,0) -- (-0.3,0);
		\node at (0,-.4) {$r_0^{\rm IR}$};
	}\, 
	{S}^{d'}_{(X,P)(Y,\cdot)}\left({\tiny
\begin{array}{c}
\{[N]_{d'},-N\}\\
\{[N-1]_{d'},-N+1\}\\
\vdots\\
\{[N-d'+1]_{d'},-N+d'-1\}
\end{array}}
\right)
\end{equation*}
One can think of the matrix factorization associated to $R^{\rm IR}$ as the limit $\lim_{N\to\infty} R_N^{\rm IR}$. Note that $N$ only shifts the charges of this matrix factorization!

As we will see later, left-fusion with the defect $R^{\rm IR}$ just sets $X$ to $1$ in the matrix factorization $R^{\rm IR}$ is fused with.

In the following we will also need the defects $R^{\rm UV}$ obtained by pushing the GLSM identity defect to the UV on the left. The associated module is given by
\begin{equation*}
M_{I}^{\rm GLSM\,UV}={C}_{(X,\cdot)(Y,Q)}^{(\alpha,\alpha^{-1})}/
((X-\alpha^{d'}Y),(\alpha^d -Q))\,,
\end{equation*}
which can be obtained as a limit $N\to-\infty$ of the truncated modules\footnote{Note that compared to the IR case, truncation is implemented in the opposite direction.}
\begin{equation*}
M_{I}^{\rm GLSM\,UV}(N)=
\alpha^N{C}_{(X,\cdot)(Y,Q)}[\alpha]/
((X-\alpha^{d'}Y),(Q -\alpha^{d}))\alpha^N{C}_{(X,\cdot)(Y,Q)}[\alpha]\,.
\end{equation*}
An analysis analogous to the IR case yields an associated matrix factorization
\begin{equation*}
R^{\rm UV}_N:
{S}^{d}_{(X,P)(Y,\cdot)}\left({\tiny
\begin{array}{c}
\{[N+d']_{d},-N\}\\
\{[N+1+d']_{d},-N-1\}\\
\vdots\\
\{[N+d-1+d']_{d},-N-d+1\}
\end{array}}
\right)
\\ 
	\tikz[baseline=0]{
		\node at (0,.5) {$r_1^{\rm UV}$};
		\draw[arrow position = 1] (-0.3,.2) -- (0.3,.2);
		\draw[arrow position = 1] (0.3,0) -- (-0.3,0);
		\node at (0,-.4) {$r_0^{\rm UV}$};
	}\, 
	{S}^{d}_{(X,P)(Y,\cdot)}\left({\tiny
\begin{array}{c}
\{[N]_{d},-N\}\\
\{[N+1]_{d},-N-1\}\\
\vdots\\
\{[N+d-1]_{d},-N-d+1\}
\end{array}}
\right)
\end{equation*}
with $r_1^{\rm UV}=(X\bI_d-Y\epsilon_d^{d'}I_Q)$, $r_0^{\rm UV}=\prod_{i=1}^{d-1}=(X\bI_d-\xi^iY\epsilon_d^{d'}I_Q)$, where $I_Q$ is the diagonal matrix with $1$ in its first $d-d'$ diagonal entries and $Q$ in its last $d'$.
Explicitly,
\begin{equation*}
r_1^{\rm UV}=\left(\begin{array}{cccccc}
X&&&-QY&&\\
&\ddots&&&\ddots&\\
&&\ddots&&&-QY\\
-Y&&&\ddots&&\\
&\ddots&&&\ddots&\\
&&-Y&&&X
\end{array}\right)\,.
\end{equation*}
The matrix factorization $R^{\rm UV}$ can then be thought of as the limit $\lim_{N\to-\infty} R^{\rm UV}_N$.
Left-fusion with $R^{\rm UV}$ implements the push-down to the UV phase, i.e.~setting $P$ to $1$.

\subsection{Projection defects}
In the previous section we have shown that the defects $RG_N$ describing the transition from UV to IR phase factorize as $RG_N\cong R^{\rm IR}\otimes T_N^{\rm UV}$. Here $T_N^{\rm UV}$ is the defect lifting the UV phase into the GLSM, and $R^{\rm IR}$ is the defect from the GLSM to the IR phase implementing the push-down to the IR.

Indeed, we can also consider the fusion $R^{\rm UV}\otimes T_N^{\rm UV}$. This defect describes the lift of the UV to the GLSM and the subsequent push-down to the same phase. 
Since identity defects are idempotent, $I\otimes I\cong I$, this defect can be obtained by pushing down to the UV phase on both sides of the GLSM identity defect, combined with a truncation. The untruncated push-down was calculated in section~\ref{sec:glsmidtolgid}. 
The result is the identity defect of the UV phase. Indeed, it is not difficult so see that the truncation essentially does not change the calculation, and that also the truncated push-down yields the identity defect of the UV phase
\begin{equation}\label{eq:rtimest}
R^{\rm UV}\otimes T_N^{\rm UV}\cong I^{\rm UV}\,.
\end{equation}
Another way to obtain this result is to use the fact (discussed below) that left-fusion with defects $R^i$ just implements the push-down to phase$_i$, i.e.~it just sets the variable to $1$, which is associated to the field 
obtaining a non-trivial vacuum expectation value in  phase$_i$.
In the case of $R^{\rm UV}$, this is the variable $P$. Setting
$P=1$ in the matrix factorization $T_N^{\rm UV}$ given in (\ref{eq:tnuv})
indeed yields the UV identity matrix factorization $I^{\rm UV}$, c.f.~(\ref{eq:uvid}).

Of course, one can also straight-forwardly calculate the fusion. As we already used in section~\ref{sec:identity},
on the level of modules, fusion corresponds to the part of the tensor product which is invariant under the gauge group of the model in between the fused defects \cite{Brunner:2007ur}.
One obtains
\begin{eqnarray}
&&
\left(M_I^{\rm GLSM\,UV}\otimes M_I^{\rm UV\,GLSM}(N)\right)^{U(1)}\nonumber\\
&&\quad\cong
\left(
{{C}^{(\alpha,\alpha^{-1})}_{(X,\cdot)(Y,Q)}\over((X-\alpha^{d'}Y),(\alpha^d-Q))}
\otimes_{\bC[Y,Q]}{
\beta^NC_{(Y,Q)(Z,\cdot)}[\beta^{-1}]\over(
(Z-\beta^{-d'}Y),(Q-\beta^{-d})
)
\beta^NC_{(Y,Q)(Z,\cdot)}[\beta^{-1}]}
\right)^{U(1)}\nonumber\\
&&\quad\cong
\left({
\beta^NC_{(X,\cdot)(Z,\cdot)}[\alpha,\alpha^{-1},\beta^{-1}]\over
(
(\alpha\alpha^{-1}-1),(Z-(\alpha\beta)^{-d'}X),(\alpha^d-\beta^{-d})
)
\beta^NC_{(X,\cdot)(Z,\cdot)}[\alpha,\alpha^{-1},\beta^{-1}]}
\right)^{U(1)}\nonumber\\
&&\quad\cong
{
(\alpha\beta)^NC_{(X,\cdot)(Z,\cdot)}[(\alpha\beta)^{-1}]\over
(
(Z-(\alpha\beta)^{-d'}X),(1-(\alpha\beta)^{-d})
)
(\alpha\beta)^NC_{(X,\cdot)(Z,\cdot)}[(\alpha\beta)^{-1}]}
\,.\nonumber
\end{eqnarray}
With the same arguments as in section~\ref{sec:glsmidtolgid} this can be seen to be a module associated to the identity matrix factorization $I^{\rm UV}$ of the Landau-Ginzburg orbifold in the UV.

Analogously one finds
\begin{equation*}
R^{\rm IR}\otimes T_N^{\rm IR}\cong {I}^{\rm IR}\,.
\end{equation*}

Relation (\ref{eq:rtimest}) implies that the defect
\begin{equation*}
P_N^{\rm UV}=T_N^{\rm UV}\otimes R^{\rm UV}
\end{equation*}
is idempotent, i.e.~$P_N^{\rm UV}\otimes P_N^{\rm UV}\cong P_N^{\rm UV}$. 
This defect realizes the UV phase inside the GLSM in the sense of \cite{Klos:2019axh}. In particular, the category of D-branes in the UV phase is equivalent to the subcategory of GLSM branes invariant under fusion with $P_N^{\rm UV}$.

A module associated to $P_N^{\rm UV}$ can be obtained as
\begin{eqnarray}
M_{P_N^{\rm UV}}&=&M_{I}^{\rm UV\,GLSM}(N)\otimes M_{I}^{\rm GLSM\,UV}\nonumber\\
&=&\left(
{\alpha^NC_{(X,P)(Y,\cdot)}[\alpha^{-1}]\over
(
(Y-\alpha^{-d'}X),(P-\alpha^{-d})
)
\alpha^NC_{(X,P)(Y,\cdot)}[\alpha^{-1}]}
\otimes_{\bC[Y]}
{{C}^{(\beta,\beta^{-1})}_{(Y,\cdot)(Z,R)}\over
((Y-\beta^{d'}Z),(\beta^d-R))}
\right)^{\bZ_d}\nonumber\\
&\cong&
\left(
{\alpha^NC_{(X,P)(Z,R)}[\alpha^{-1},\beta,\beta^{-1}]\over
((\beta\beta^{-1}-1),(P-\alpha^{-d}),(\beta^d-R),(\alpha^{-d'}X-\beta^{d'}Z)
)\alpha^NC_{(X,P)(Z,R)}[\alpha^{-1},\beta,\beta^{-1}]}
\right)^{\bZ_d}\nonumber
\end{eqnarray}
The $\bZ_d$-invariant generators are given by $(\alpha\beta)^{N-i}(\beta^d)^m$ for $i\in\mathbb{N}_0$ and $m\in\bZ$. They carry $U(1)\times U(1)$-charges $(N-i,-N+i-md)$. Defining $\gamma:=\alpha\beta$ and $\delta:=\beta^{-d}$ of $U(1)\times U(1)$ charges $(1,-1)$ and $(0,d)$ respectively, one can write
\begin{equation*}
M_{P_N^{\rm UV}}\cong
{\gamma^NC_{(X,P)(Z,R)}[\gamma^{-1},\delta]\over
((P-\gamma^{-d} R),(X\gamma^{-d'}-Z),(\delta R-1))
\gamma^NC_{(X,P)(Z,R)}[\gamma^{-1},\delta]}\,.
\end{equation*}
Note that $R$ is invertible in this module! It can be considered as a module over the ring 
\begin{equation*}
C_{(X,P)(Z,R)}^{\delta}:={C_{(X,P)(Z,R)}[\delta]\over
(\delta R-1)C_{(X,P)(Z,R)}[\delta]
}
\end{equation*}
in which $R$ is invertible. Over this ring, $M_{P_N^{\rm UV}}$ is finitely generated with generators $e_i:=\gamma^{N-d+1+i}$, $0\leq i<d$ of $U(1)\times U(1)$-charges $(N-d+1+i,-N+d-1-i)$. They satisfy relations
\begin{eqnarray}
Xe_i=
Ze_{i+d'}\,,&&\quad i+d'<d\nonumber\\
PXe_i=RZe_{i+d'-d}\,,&&\quad i+d'\geq d\nonumber
\end{eqnarray}
Thus, as module over the ring $C_{(X,P)(Z,R)}^{\delta}$, $M_{P_N^{\rm UV}}$ is isomorphic to the cokernel of the map
\begin{equation*}
p_1:\left(C_{(X,P)(Z,R)}^{\delta}\right)^d
\left({\tiny
\begin{array}{c}
\{N-d+1+d',-N+d-1\}\\
\{N-d+2+d',-N+d-2\}\\
\vdots\\
\{N,-N+d'\}\\
\{N-d+1,-N+d'-1\}\\
\{N-d+2,-N+d'-2\}\\
\vdots\\
\{N-d+d',-N\}
\end{array}}
\right)
\rightarrow
(C^{\delta}_{(X,P)(Z,R)})^d
\left({\tiny
\begin{array}{c}
\{N-d+1,-N+d-1\}\\
\{N-d+2,-N+d-2\}\\
\vdots\\
\{N,-N\}
\end{array}}
\right)\,,
\end{equation*}
with $p_1=(XI_P-Z\epsilon_d^{d'}I_R)$. Here $I_P$ is the diagonal $d\times d$-matrix whose first $d-d'$ diagonal entries are $1$ and whose last $d'$ diagonal entries are $P$, and $I_R$ is the diagonal $d\times d$-matrix whose first $d-d'$ entries are $1$ and whose last $d'$ entries are $R$. 
Concretely,
\begin{equation*}
p_1=\left(\begin{array}{cccccccc}
X&&&-PZ&&&&\\
&\ddots&&&\ddots&&&\\
&&X&&&\ddots&&\\
&&&PX&&&\ddots&\\
&&&&\ddots&&&-PZ\\
-Z&&&&&\ddots&&\\
&\ddots&&&&&\ddots&\\
&&-Z&&&&&PX
\end{array}
\right)
\end{equation*}
Note that\footnote{
In fact, this is true for any choice of diagonal matrices $I_P=\text{diag}(P^{n_1},\ldots,P^{n_d})$ and $I_R=\text{diag}(R^{m_1},\ldots,R^{m_d})$ with $\sum n_i=d'=\sum m_i$.}
\begin{equation*}
\prod_{i=0}^{d-1}\left(X
\epsilon_d^{-id'}I_P\epsilon_d^{id'} -\xi^iZ\epsilon_d^{d'}I_R
\right)=X^dP^{d'}-Z^dR^{d'}\,.
\end{equation*}
Hence, $p_1$ together with $p_0=\prod_{i=1}^{d-1}\left(X
\epsilon_d^{-id'}I_P\epsilon_d^{id'} -\xi^iZ\epsilon_d^{d'}I_R
\right)$ forms a matrix factorization 
\begin{equation*}
P^{\rm UV}_N:
\left({S}^{\delta}_{(X,P)(Z,R)}\right)^d\left({\tiny
\begin{array}{c}
\{N-d+1+d',-N+d-1\}\\
\{N-d+2+d',-N+d-2\}\\
\vdots\\
\{N,-N+d'\}\\
\{N-d+1,-N+d'-1\}\\
\{N-d+2,-N+d'-2\}\\
\vdots\\
\{N-d+d',-N\}
\end{array}}
\right)
\\ 
	\tikz[baseline=0]{
		\node at (0,.5) {$p_1$};
		\draw[arrow position = 1] (-0.3,.2) -- (0.3,.2);
		\draw[arrow position = 1] (0.3,0) -- (-0.3,0);
		\node at (0,-.4) {$p_0$};
	}\, 
	\left({S}^{\delta}_{(X,P)(Z,R)}\right)^d\left({\tiny
\begin{array}{c}
\{N-d+1,-N+d-1\}\\
\{N-d+2,-N+d-2\}\\
\vdots\\
\{N,-N\}
\end{array}}
\right)
\end{equation*}
of $W(X,P)-W(Y,Q)$ over the ring $S_{(X,P)(Z,R)}^{\delta}=\bC[X,P,Z,R,\delta]/(\delta R-1)\bC[X,P,Z,R,\delta]$ of chiral fields of the GLSM on the left and  right of the defect, in which the field $R$ is made invertible. 

\subsection{Action on D-branes}

Here, we will discuss the action of the defects $T_N^i$, $R^i$ and $P_N^i$ on D-branes (boundary conditions).

\subsubsection{\texorpdfstring{$R^i$}{Ri}}

Fusion with a defect $R^i$ acts on D-branes by pushing down the respective GLSM matrix factorizations to 
 phase$_i$ by setting the variable obtaining a vacuum expectation value in the phase to $1$. More precisely, let 
\begin{equation*}
{P}: 
P_1=S_{(Y,Q)}^r\left\{\begin{array}{c}b_1\\\vdots\\b_r\end{array}\right\}
\\ 
	\tikz[baseline=0]{
		\node at (0,.5) {$
		{p}_1$};
		\draw[arrow position = 1] (-0.3,.2) -- (0.3,.2);
		\draw[arrow position = 1] (0.3,0) -- (-0.3,0);
		\node at (0,-.4) {${p}_0$};
	}\, 
	S_{(Y,Q)}^r\left\{\begin{array}{c}a_1\\\vdots\\a_r\end{array}\right\}=P_0
	\end{equation*}
be a $U(1)$-equivariant matrix factorization of $Y^dQ^{d'}$ representing a  D-brane in the GLSM. 
Here, we use the following notation:
\begin{eqnarray}
S_{(Y,Q)}&=&\bC[Y,Q]\nonumber\\
C_{(Y,Q)}&=&S_{(Y,Q)}/(Y^dQ^{d})\nonumber\,.
\end{eqnarray}
As before, replacing one of the variables in the subscript with a `$\cdot$' means that we set the respective variable to $1$. So, in particular $S_{(Y,\cdot)}=\bC[Y]$ and $C_{(Y,\cdot)}=\bC[Y]/(Y^d)$. To this matrix factorization we associate the $C_{(X,P)}$-module
$$
M_P=\text{coker}\left(
P_1':=
C_{(Y,Q)}^r\left\{\begin{array}{c}b_1\\\vdots\\b_r\end{array}\right\}
\stackrel{p_1}{\longrightarrow}
C_{(Y,Q)}^r\left\{\begin{array}{c}a_1\\\vdots\\a_r\end{array}\right\}=:P_0'
\right)\,.
$$
We can now calculate the fusion
$\widehat{P}=R^{\rm UV}\otimes P$. On the level of modules we obtain an associated 
$C_{(X,\cdot)}$-module 
\begin{eqnarray}
M_{\widehat{P}}&=&
\left(M_I^{\rm GLSM\,UV}\otimes_{S_{(Y,Q)}}M_P\right)^{U(1)}\nonumber\\
&=&
\left({{C}^{(\alpha,\alpha^{-1})}_{(X,\cdot),(Y,Q)}\over
((Y-\alpha^{-d'}X),(Q-\alpha^{d}))}
\otimes_{S_{(Y,Q)}}M_P\right)^{U(1)}\nonumber
\end{eqnarray}
The $U(1)$-invariant generators of this module are given $\hat{e}_i:=\alpha^{a_i}e_i$, where the $e_i$, $1\leq i \leq r$ are the generators the module $P'_0$ of $U(1)$-charge $a_i$. Note that $\alpha$ has $\bZ_d\times U(1)$-charge $([1]_d,-1)$, and hence, the $\bZ_d$-charge of $\hat{e}_i$ are just the induced $\bZ_d$-charges $[a_i]_d$. The relations from the first tensor factor set the variable $Y$ to $\alpha^{-d'}X$ and $Q$ to $\alpha^{d}$. The relations from the second tensor factor, coming from the matrix $p_1$ can then be written in terms of the matrix $\widehat{p}_1=p_1(Y=X,Q=1)$ obtained from $p_1$ by setting $Y$ to $X$ and $Q$ to $1$. One obtains
\begin{equation*}
M_{\widehat{P}}\cong\text{coker}
\left(
\widehat{P}_1':=
C^r_{(X,\cdot)}\left\{\begin{array}{c} {[b_1]_d}\\\vdots\\ {[b_r]_d}\end{array}\right\}
\stackrel{\widehat{p}_1}{\longrightarrow}
C^r_{(X,\cdot)}\left\{\begin{array}{c} {[a_1]_d}\\\vdots\\ {[a_r]_d}\end{array}\right\}
=:\widehat{P}_0'
\right)\,.
\end{equation*}
This module is associated to the $\bZ_d$-equivariant matrix factorization
\begin{equation*}
\widehat{P}: 
\widehat{P}_1=S_{(X,\cdot)}^r\left\{\begin{array}{c}{[b_1]_d}\\\vdots\\{[b_r]_d}\end{array}\right\}
\\ 
	\tikz[baseline=0]{
		\node at (0,.5) {$
		\widehat{p}_1$};
		\draw[arrow position = 1] (-0.3,.2) -- (0.3,.2);
		\draw[arrow position = 1] (0.3,0) -- (-0.3,0);
		\node at (0,-.4) {$\widehat{p}_0$};
	}\, 
	S_{(X,\cdot)}^r\left\{\begin{array}{c}{[a_1]_d}\\\vdots\\{[a_r]_d}\end{array}\right\}=\widehat{P}_0
	\end{equation*}
of $X^d$. The matrices $\widehat{p}_i$ are obtained from the respective $p_i$ by setting $Y$ to $X$ and $Q$ to $1$. 
An  analogous result holds for the action of $R^{\rm IR}$. 

Thus, $R^i$ indeed fuses with GLSM branes by setting the variables acquiring a non-trivial vacuum expectation value in phase$_i$ to $1$ in the respective matrix factorization, and breaking the gauge symmetry accordingly.

\subsubsection{\texorpdfstring{$T_N^i$}{TNi}}

Fusion with $T_N^{\rm UV}$ lifts 
 $\bZ_d$-equivariant matrix factorizations of $X^d$ to $U(1)$-equivariant matrix factorizations of $P^{d'}X^d$. Since $R^{\rm UV}\otimes T_N^{\rm UV}\cong I^{\rm UV}$ and $R^{\rm UV}$ acts by setting $P=1$,  the lifted matrix factorization has to reduce to the original one upon setting $P=1$. 
Thus, such lifts are obtained by inserting $P$'s into the matrices of the original matrix factorizations in such a way that the $\bZ_d$-representations on the matrix factorizations lift to $U(1)$-representations. In fact, for a given matrix factorization there are many possible lifts. As  it turns out, fusion with $T_N^{\rm UV}$ produces lifts whose $U(1)$-representations have charges in $\{N-d+1,N-d+2,\ldots, N\}$. 

Let us illustrate this in the example of $\bZ_d$-equivariant linear rank-$1$ factorizations
\begin{equation*}
{L}^{\rm UV}_{[a]_d}: 
S_{(Y,\cdot)}\{[a+d']_d\}
\\ 
	\tikz[baseline=0]{
		\node at (0,.5) {$Y$};
		\draw[arrow position = 1] (-0.3,.2) -- (0.3,.2);
		\draw[arrow position = 1] (0.3,0) -- (-0.3,0);
		\node at (0,-.4) {$Y^{d-1}$};
	}\, 
	S_{(Y,\cdot)}\{[a]_d\}
	\end{equation*}
of $X^d$.
These matrix factorizations 
generate the category of $\bZ_d$-equivariant matrix factorizations of $X^d$, i.e. the category of UV D-branes. 

Now, any of the $U(1)$-equivariant rank-$1$ matrix factorizations
\begin{equation*}
{L}^{\rm GLSM}_{c,m}: 
S_{(X,P)}\{c+d' -md\}
\\ 
	\tikz[baseline=0]{
		\node at (0,.5) {$YQ^m$};
		\draw[arrow position = 1] (-0.3,.2) -- (0.3,.2);
		\draw[arrow position = 1] (0.3,0) -- (-0.3,0);
		\node at (0,-.4) {$Y^{d-1}Q^{d'-m}$};
	}\, 
	S_{(Y,Q)}\{c\}
	\end{equation*}
of $X^dP^{d'}$ is a lift of $L^{\rm UV}_{[a]_d}$ for $c\in a+d\bZ$ and $0\leq m\leq d'$. Namely,
$$
R^{\rm UV}\otimes {L}^{\rm GLSM}_{c,m}\cong L^{\rm UV}_{[a]_d}\,,
$$
or to put it differently, 
setting $Y=X$ and $Q=1$ in ${L}^{\rm GLSM}_{c,m}$ produces ${L}^{\rm UV}_{[a]_d}$.

Next, we will compute to which of the lifts ${L}^{\rm GLSM}_{c,m}$ a matrix factorization 
${L}^{\rm UV}_{[a]_d}$ is mapped under fusion with $T_N^{\rm UV}$. As before we will compute the fusion on the level of modules. To ${L}^{\rm UV}_{[a]_d}$ we associate the $C_{(Y,\cdot)}$-module
\begin{equation*}
M_{{L}_{[a]_d}^{\rm UV}}=C_{(Y,\cdot)}\{[a]_d\}/(Y)\,.
\end{equation*}

The fusion $T_N^{\rm UV}\otimes {L}^{\rm UV}_{[a]_d}$ is then given by the matrix factorization associated to the 
$C_{(X,P)}$-module given by the $\bZ_d$-invariant part of the tensor product
\begin{eqnarray}
&&\left(M_{I}^{\rm UV\,GLSM}\otimes_{\bC[Y]}M_{L_a^{\rm UV}}\right)^{\bZ_d}\nonumber\\
&&\qquad\qquad=
\left({\alpha^{N}C_{(X,P)(Y,\cdot)}[\alpha^{-1}]\over
((Y-\alpha^{-d'}X),(P-\alpha^{-d}))\alpha^N{C}_{(X,P)(Y,\cdot)}[\alpha^{-1}]}
\otimes_{\bC[Y]}{C_{(Y,\cdot)}\{[a]_d\}\over (Y)}\right)^{\bZ_d}\nonumber\\
&&\qquad\qquad\cong
\left(
{\alpha^NC_{(X,P)}[\alpha^{-1}]\{a\}\over
((P-\alpha^{-d}),\alpha^{-d'}X)\alpha^NC_{(X,P)}[\alpha^{-1}]\{a\}}
\right)^{\bZ_d}\nonumber \ .
\end{eqnarray}
There is just one $\bZ_d$-equivariant generator of this module over $C_{(X,P)}$, namely
$\alpha^{N-\{N-a\}_d}$ of $U(1)$-charge $N-\{N-a\}_d$. Here $\{\cdot\}_d$ denotes the representative of the rest class $[\cdot]_d$ modulo $d$ in the range $\{0,\ldots, d-1\}$. There is one relation, namely 
\begin{equation} \label{eq:nselect}
P^nX\alpha^{N-\{N-a\}_d}=0\,,\quad\text{where}\quad n=\left\{\begin{array}{ll}0\,,&d'-\{N-a\}_d\leq 0\\
1\,,&d'-\{N-a\}_d>0\end{array}\right. \ .
\end{equation}
Hence,
\begin{equation*}
\left(M_{I}^{\rm UV\,GLSM}\otimes_{\bC[Y]}M_{{L}_{[a]_d}^{\rm UV}}\right)^{\bZ_d}
\cong C_{(X,P)}/P^nXC_{(X,P)}\,.\nonumber
\end{equation*}
which is associated to the matrix factorization
\begin{equation*}
S_{(X,P)}\{N-\{N-a\}_d+d'-nd\}
\\ 
	\tikz[baseline=0]{
		\node at (0,.5) {$XP^n$};
		\draw[arrow position = 1] (-0.3,.2) -- (0.3,.2);
		\draw[arrow position = 1] (0.3,0) -- (-0.3,0);
		\node at (0,-.4) {$X^{d-1}P^{d'-n}$};
	}\, 
	S_{(X,P)}\{N-\{N-a\}_d\} \ ,
	\end{equation*}
This is nothing but
${L}^{\rm GLSM}_{N-\{N-d\}_d,n}$, where the value of $n$ depends on $a$ as stated in
(\ref{eq:nselect}). Hence:
$$
T_N^{\rm UV}\otimes {L}^{\rm UV}_{[a]_d}
\cong {L}^{\rm GLSM}_{N-\{N-d\}_d,n}\,.
$$
Note that due to the specific dependence of $n$ on $a$, the $U(1)$-charges of the generators (of the module of) the matrix factorization lie in the set
$\{N-d+1,N-d+2,\ldots,N\}$ of $d$ consecutive integers $\leq N$. 

Indeed, this is the way $T_N^{\rm UV}$ acts on any boundary condition\footnote{or more generally defects}. It lifts the $\bZ_d$-equivariant matrix factorization of $X^d$ to a $U(1)$-equivariant matrix factorization of $P^{d'}X^d$ by inserting factors of $P$ into the matrix factorization in such a way that the $\bZ_d$-representation lifts to $U(1)$, and that furthermore the $U(1)$-charges of the lifted representation all lie in 
$\{N-d+1,N-d+2,\ldots,N\}$. More precisely, let
\begin{equation*}
P: 
S_{(Y,\cdot)}^r \left\{\begin{array}{c} {[b_1]_d} \\ \vdots\\ {[b_r]_d}\end{array}\right\}
\\ 
	\tikz[baseline=0]{
		\node at (0,.5) {$p_1$};
		\draw[arrow position = 1] (-0.3,.2) -- (0.3,.2);
		\draw[arrow position = 1] (0.3,0) -- (-0.3,0);
		\node at (0,-.4) {$p_0$};
	}\, 
	S_{(Y,\cdot)}^r\left\{\begin{array}{c} {[a_1]_d}\\ \vdots\\ {[a_r]_d}\end{array}\right\}
	\end{equation*}
be a rank-$r$ $\bZ_d$-equivariant matrix factorization of $Y^d$. Then one can show that 
$T_N^{\rm UV}\otimes P$ is given by the $U(1)$-equivariant matrix factorization
\begin{equation*}
\widehat{P}: 
S_{(X,P)}^r\left\{\begin{array}{c}N-\{N-b_1\}_d\\\vdots\\N-\{N-b_r\}_d\end{array}\right\}
\\ 
	\tikz[baseline=0]{
		\node at (0,.5) {$
		\widehat{p}_1$};
		\draw[arrow position = 1] (-0.3,.2) -- (0.3,.2);
		\draw[arrow position = 1] (0.3,0) -- (-0.3,0);
		\node at (0,-.4) {$\widehat{p}_0$};
	}\, 
	S_{(X,P)}^r\left\{\begin{array}{c}N-\{N-a_1\}_d\\\vdots\\N-\{N-a_r\}_d\end{array}\right\}
	\end{equation*}
of $X^dP^{d'}$, where the matrix $\widehat{p}_1$ is obtained from $p_1$ by 
replacing each monomial $Y^r$ in the matrix entry $(p_1)_{ij}$ by $P^nX^r$,
with
\begin{equation*}
n=\text{max}\{0,-\left(\{N-a_i\}_d-d'r\right)\text{div }d\}\,.
\end{equation*}
`$\text{div}$' denotes the division with (non-negative) remainder. Similarly $\widehat{p}_0$ is obtained from $p_0$ by replacing monomials $Y^r$ in $(p_0)_{ij}$ by $P^nX^r$ with 
\begin{equation*}
n=\text{max}\{0,-\left(\{N-b_i\}_d-d'r\right)\text{div }d\}\,.
\end{equation*}
One arrives at a similar conclusion for the action of $T_N^{\rm IR}$, where however the $U(1)$-charges of the lifted matrix factorization have to lie in the smalle set $\{N-d'+1,\ldots,N\}$ of $d'$ consecutive integers $\leq N$.

\subsubsection{\texorpdfstring{$P_N^i$}{PNi}}

 Since fusion is associative, the last two sections imply the following action of the projection defects $P_N^{\rm UV}\cong T_N^{\rm UV}\otimes R^{\rm UV}$.
Let 
\begin{equation*}
P: 
S_{(Y,Q)}^r \left\{\begin{array}{c} {b_1} \\ \vdots\\ {b_r}\end{array}\right\}
\\ 
	\tikz[baseline=0]{
		\node at (0,.5) {$p_1$};
		\draw[arrow position = 1] (-0.3,.2) -- (0.3,.2);
		\draw[arrow position = 1] (0.3,0) -- (-0.3,0);
		\node at (0,-.4) {$p_0$};
	}\, 
	S_{(Y,Q)}^r\left\{\begin{array}{c} {a_1}\\ \vdots\\ {a_r}\end{array}\right\}
	\end{equation*}
be a $U(1)$-equivariant matrix factorization of $Y^dQ^{d'}$. Then $P_N^{\rm UV}\otimes P$ is isomorphic to the $U(1)$-equivariant matrix factorization
\begin{equation*}
\widehat{P}: 
S_{(X,P)}^r\left\{\begin{array}{c}N-\{N-b_1\}_d\\\vdots\\N-\{N-b_r\}_d\end{array}\right\}
\\ 
	\tikz[baseline=0]{
		\node at (0,.5) {$
		\widehat{p}_1$};
		\draw[arrow position = 1] (-0.3,.2) -- (0.3,.2);
		\draw[arrow position = 1] (0.3,0) -- (-0.3,0);
		\node at (0,-.4) {$\widehat{p}_0$};
	}\, 
	S_{(X,P)}^r\left\{\begin{array}{c}N-\{N-a_1\}_d\\\vdots\\N-\{N-a_r\}_d\end{array}\right\}
	\end{equation*}
of $X^dP^{d'}$. Here the matrix $\widehat{p}_1$ is obtained from $p_1$ by 
replacing each monomial $Y^rQ^s$ in the matrix entry $(p_1)_{ij}$ by $X^rP^n$,
with 
\begin{equation*}
n=\text{max}\{0,-\left(\{N-a_i\}_d-d'r\right)\text{div }d\}\,.
\end{equation*}
Similarly $\widehat{p}_0$ is obtained from $p_0$ by replacing monomials $Y^rQ^s$ in $(p_0)_{ij}$ by $X^rP^n$ with 
\begin{equation*}
n=\text{max}\{0,-\left(\{N-b_i\}_d-d'r\right)\text{div }d\}\,.
\end{equation*}
Thus, the matrix factorization $\widehat{P}$ is obtained from $P$ by shifting all $U(1)$-charges into the range $\{N-d+1,\ldots,N\}$ by adding integer multiples of $d$, setting all $Q$ in the matrices to $1$ and inserting factors of $P$ in a way ensuring $U(1)$-equivariance of $\widehat{P}$.

One finds an analogous result for $P^{\rm IR}$, where the charges are shifted by integer multiples of $d'$ into the smaller set $\{N-d'+1,\ldots,N\}$, 
$Y$ is set to $1$ and factors of $X$ are inserted in a way ensuring $U(1)$-equivariance.

\subsubsection{\texorpdfstring{$RG_N$}{RGN}}
As alluded to above,
the defects $RG_N$ describing the transitions between UV and IR phase are special RG defects between the Landau-Ginzburg orbifolds in the UV and the IR. The action of general RG defects have been discussed at length in \cite{Brunner:2007ur}. In particular, there is an instructive picture of the D-brane transport coming from the corresponding flow between unorbifolded Landau-Ginzburg models in the mirror theory. These flows are tiggered by lower order perturbations of the superpotential $W(X)=X^d+\sum_{i<d}\lambda_iX^i$. During the flows some vacua of the theory, corresponding to critical points of $W$ move off to infinity and decouple, taking with them certain A-branes attached to them. (For more details see \cite{Brunner:2007ur}.)

The factorization $RG_N\cong R^{\rm IR}\otimes T_N^{\rm UV}$ together with the action of the $R^{\rm IR}$ and
$T_N^{\rm UV}$ discussed in the previous sections now leads to a stepwise description of the action of $RG_N$. 
Start with a D-brane in the UV phase. For simplicity we only discuss D-branes described by a rank-$1$ matrix factorizations
\begin{equation}\label{eq:startmf}
P: S_{(Y,\cdot)}\{[a+rd']_d\}
\\ 
	\tikz[baseline=0]{
		\node at (0,.5) {$Y^r$};
		\draw[arrow position = 1] (-0.3,.2) -- (0.3,.2);
		\draw[arrow position = 1] (0.3,0) -- (-0.3,0);
		\node at (0,-.4) {$Y^{d-r}$};
	}\, 
	S_{(Y,\cdot)}\{[a]_d\}\,.
	\end{equation}
Under the action of $T_N^{\rm UV}$ $P$ gets lifted to the $U(1)$-equivariant matrix factorization
\begin{equation*}
P': S_{(X,P)}\{N-\{N-a-rd'\}_d\}\}
\\ 
	\tikz[baseline=0]{
		\node at (0,.5) {$X^rP^n$};
		\draw[arrow position = 1] (-0.3,.2) -- (0.3,.2);
		\draw[arrow position = 1] (0.3,0) -- (-0.3,0);
		\node at (0,-.4) {$X^{d-r}P^{d'-n}$};
	}\, 
	S_{(X,P)}\{N-\{N-a\}_d\}\,,
	\end{equation*}
	of $X^dP^{d'}$, 
	where 
\begin{equation}\label{eq:nlinear}
	nd=rd'+\{N-a-rd'\}_d-\{N-a\}_d\quad
	\Longrightarrow\quad n= rd'\,\text{div}\,d+\left\{\begin{array}{ll}0\,,&\{N-a\}_d\geq\{rd'\}_d\\1\,,&\{N-a\}_d<\{rd'\}_d\end{array}\right.
\end{equation}
$R^{\rm IR}$ then pushes down this matrix factorization to the IR Landau-Ginzburg model by setting $X=1$, resulting in the $\bZ_{d'}$-equivariant matrix factorization
\begin{equation}\label{eq:p''}
P'': S_{(\cdot,P)}\{[N-\{N-a-rd'\}_d\}]_{d'}\}
\\ 
	\tikz[baseline=0]{
		\node at (0,.5) {$P^n$};
		\draw[arrow position = 1] (-0.3,.2) -- (0.3,.2);
		\draw[arrow position = 1] (0.3,0) -- (-0.3,0);
		\node at (0,-.4) {$P^{d'-n}$};
	}\, 
	S_{(\cdot,P)}\{[N-\{N-a\}_d]_{d'}\}\,,
	\end{equation}
of $P^{d'}$. Thus, $RG_N\otimes P\cong P''$. 

Note that in case $n=0$ and $n=d'$ in (\ref{eq:nlinear}), the matrix factorization $P''$ is trivial, and hence the D-brane corresponding to the matrix factorization $P$ in (\ref{eq:startmf}) decouples under the RG flow. That is the case whenever $rd'<d$ and $\{N-a\}_d\geq\{rd'\}_d$ ($n=0$) or $r\geq d-s$ with $sd'\leq d$ and $\{N-a\}_d<\{-sd'\}_d$ ($n=d'$).

So for instance, the degree-$1$ linear matrix factorizations $P$ (i.e. those with $r=1$) with 
$a=N-b$ for $d>b\geq d'$ decouple under the RG flow, whereas the ones for $d'>b\geq 0$ are mapped to degree-$1$ linear matrix factorizations in the IR. 

In general, it follows from (\ref{eq:nlinear}) that $n\leq r$, so the 
degree of the matrix factorization does not increase during the flow. Either it stays the same, or it decreases.
A decrease means that the corresponding D-brane decays during the flow and at least one constituent decouples.

Let us illustrate this in a specific example, namely for $d=8$ and $d'=5$, i.e. we are considering a $U(1)$-GLSM with superpotential $W=X^8P^5$, where the $U(1)$-charges of $X$ and $P$ are $5$ and $-8$, respectively.
The transition defects $RG_N$ describe a certain RG flows between the Landau-Ginzburg orbifolds $X^8/\bZ_8$ and $P^5/\bZ_5$. For simplicity we will discuss the action of $RG_0$, i.e.~we set $N=0$.\footnote{$N$ can be shifted by a quantum symmetry. This is a charge shift which can  be implemented by a charge shifted versions  of the identity defect  in the respective LG orbifold \cite{Brunner:2007ur}.}
Let us first consider the action on linear rank-$1$ factorizations
\begin{equation}\label{eq:p1}
P: S_{(Y,\cdot)}\{[-b+5]_8\}
\\ 
	\tikz[baseline=0]{
		\node at (0,.5) {$Y$};
		\draw[arrow position = 1] (-0.3,.2) -- (0.3,.2);
		\draw[arrow position = 1] (0.3,0) -- (-0.3,0);
		\node at (0,-.4) {$Y^{7}$};
	}\, 
	S_{(Y,\cdot)}\{[-b]_8\}
\end{equation}
for $0\leq b<8$. Under the action of $T_0$ these are mapped to the matrix factorizations
\begin{equation}\label{eq:p'1}
P': S_{(X,P)}\{-b+5\}
\\ 
	\tikz[baseline=0]{
		\node at (0,.5) {$X$};
		\draw[arrow position = 1] (-0.3,.2) -- (0.3,.2);
		\draw[arrow position = 1] (0.3,0) -- (-0.3,0);
		\node at (0,-.4) {$X^{7}P^{5}$};
	}\, 
	S_{(X,P)}\{-b\}\,,
	\end{equation}
for $5\leq b<8$, and to
\begin{equation}\label{eq:p'2}
P': S_{(X,P)}\{-b-3\}
\\ 
	\tikz[baseline=0]{
		\node at (0,.5) {$XP$};
		\draw[arrow position = 1] (-0.3,.2) -- (0.3,.2);
		\draw[arrow position = 1] (0.3,0) -- (-0.3,0);
		\node at (0,-.4) {$X^{7}P^{4}$};
	}\, 
	S_{(X,P)}\{-b\}\,,
	\end{equation}
for $0\leq b<5$. These are  the lifts of the $\bZ_8$-equivariant matrix factorizations (\ref{eq:p1}) of $X^8$
to $U(1)$-equivariant matrix factorizations of $X^8P^5$ whose charges are contained in $\{-7,-6,\ldots,0\}$.
Acting with $R^{\rm IR}$ essentially sets $X=1$ and breaks the $U(1)$ to $\bZ_{5}$. In the first case,
$5\leq b<8$, the matrix factorizations (\ref{eq:p'1}) are mapped to the trivial matrix factorizations
\begin{equation*}
P'': S_{(\cdot,P)}\{-[b]_{5}\}
\\ 
	\tikz[baseline=0]{
		\node at (0,.5) {$1$};
		\draw[arrow position = 1] (-0.3,.2) -- (0.3,.2);
		\draw[arrow position = 1] (0.3,0) -- (-0.3,0);
		\node at (0,-.4) {$P^{5}$};
	}\, 
	S_{(\cdot,P)}\{-[b]_5\}\,.
	\end{equation*}
The D-branes corresponding to (\ref{eq:p1}) for $5\leq b<8$ therefore decouple under the RG flow. 
For $0\leq b<5$, on the other hand, the matrix factorizations (\ref{eq:p'2}) are mapped to the linear factorizations
\begin{equation*}
P'': S_{(\cdot,P)}\{-[b+3]_{5}\}
\\ 
	\tikz[baseline=0]{
		\node at (0,.5) {$P$};
		\draw[arrow position = 1] (-0.3,.2) -- (0.3,.2);
		\draw[arrow position = 1] (0.3,0) -- (-0.3,0);
		\node at (0,-.4) {$P^{4}$};
	}\, 
	S_{(\cdot,P)}\{-[b]_5\}\,.
	\end{equation*}
The corresponding D-branes do not decouple.

Next, let us discuss the action on quadratic matrix factorizations
\begin{equation}\label{eq:p2}
P: S_{(Y,\cdot)}\{[-b+2]_8\}
\\ 
	\tikz[baseline=0]{
		\node at (0,.5) {$Y^2$};
		\draw[arrow position = 1] (-0.3,.2) -- (0.3,.2);
		\draw[arrow position = 1] (0.3,0) -- (-0.3,0);
		\node at (0,-.4) {$Y^{7}$};
	}\, 
	S_{(Y,\cdot)}\{[-b]_8\}\,.
\end{equation}
Acting on them with $T_0$, one obtains
\begin{equation}\label{eq:p'3}
P': S_{(X,P)}\{-b+2\}
\\ 
	\tikz[baseline=0]{
		\node at (0,.5) {$X^2P$};
		\draw[arrow position = 1] (-0.3,.2) -- (0.3,.2);
		\draw[arrow position = 1] (0.3,0) -- (-0.3,0);
		\node at (0,-.4) {$X^{6}P^{4}$};
	}\, 
	S_{(X,P)}\{-b\}\,,
	\end{equation}
for $2\leq b<8$ and
\begin{equation}\label{eq:p'4}
P': S_{(X,P)}\{-b-6\}
\\ 
	\tikz[baseline=0]{
		\node at (0,.5) {$X^2P^2$};
		\draw[arrow position = 1] (-0.3,.2) -- (0.3,.2);
		\draw[arrow position = 1] (0.3,0) -- (-0.3,0);
		\node at (0,-.4) {$X^6P^3$};
	}\, 
	S_{(X,P)}\{-b\}\,,
	\end{equation}
for $0\leq b<2$. Again, the matrix factorization $P'$ is the lift of the matrix factorization $P$ in (\ref{eq:p2}) to the GLSM whose charges lie in $\{-7,\ldots,0\}$. Acting with $R^{\rm IR}$ then yields the linear matrix factorizations
\begin{equation*}
P'': S_{(\cdot,P)}\{-[b+3]_{5}\}
\\ 
	\tikz[baseline=0]{
		\node at (0,.5) {$P$};
		\draw[arrow position = 1] (-0.3,.2) -- (0.3,.2);
		\draw[arrow position = 1] (0.3,0) -- (-0.3,0);
		\node at (0,-.4) {$P^{4}$};
	}\, 
	S_{(\cdot,P)}\{-[b]_5\}\,.
	\end{equation*}
for the case $2\leq b<8$ and the quadratic matrix factorizations
\begin{equation*}
P'': S_{(\cdot,P)}\{-[b+1]_{5}\}
\\ 
	\tikz[baseline=0]{
		\node at (0,.5) {$P^2$};
		\draw[arrow position = 1] (-0.3,.2) -- (0.3,.2);
		\draw[arrow position = 1] (0.3,0) -- (-0.3,0);
		\node at (0,-.4) {$P^{3}$};
	}\, 
	S_{(\cdot,P)}\{-[b]_5\}\,.
	\end{equation*}
for $0\leq b<2$. In the latter case, a quadratic matrix factorization is mapped to a quadratic matrix factorization under the action of $RG_0$. In the case $2\leq b<8$, the degree decreases from $2$ to $1$. Indeed, this can be completely understood in terms of the linear matrix factorizations. Namely, the quadratic matrix factorizations $P$ in 
(\ref{eq:p2}) can be written as a cone of two linear matrix factorizations as in (\ref{eq:p1}), one specified by the same label $b$ and one specified by $\{b-5\}_8$. In case both of those linear matrix factorizations survive the flow, i.e.~ for $0\leq b<2$ the quadratic matrix factorization $P$ is again mapped to a quadratic matrix factorization under $RG_0$. For the other cases, $2\leq b<8$, however, one of the two linear matrix factorizations is mapped to the trivial one under $RG_0$. Under the RG flow, the quadratic matrix factorization decays into the two constituent linear factorizations and one of them decouples. Thus, the quadratic matrix factorization flows to a linear matrix factorization.

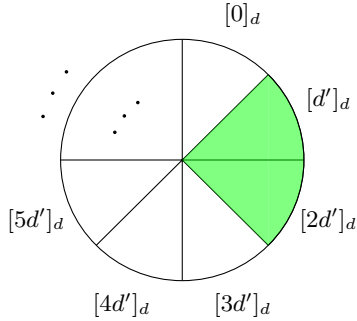
\begin{figure}[h] 
\centering
 \begin{tikzpicture}[scale=0.8]
\def\R{2cm}
\draw (0,0) circle[radius=\R];
\draw (0,0) -- (0:\R)    (0,0) -- (45:\R)     (0,0) -- (90:\R)      (0,0) -- (180:\R)    (0,0)--(225:\R)    (0,0) -- (315:\R) (0,0)--(270:\R);
\node[font=\footnotesize] at (90-22.5:\R*1.3) {$[0]_d$};
\node[font=\footnotesize] at ( 22.5:\R*1.3) {$[d']_d$};
\node[font=\footnotesize] at ( -22.5:\R*1.3) {$[2d']_d$};
\node[font=\footnotesize] at ( -90+22.5:\R*1.3) {$[3d']_d$};
\node[font=\footnotesize] at ( -135+22.5:\R*1.3) {$[4d']_d$};
\node[font=\footnotesize] at ( -180+22.5:\R*1.3) {$[5d']_d$};
\node[font=\footnotesize] at ( 270:\R*1.3) {$$};
\node[font=\footnotesize] at ( 315:\R*1.3) {$$};
\filldraw (180-22.5:\R*.6) circle[radius=0.02];
\filldraw (180-22.5-15:\R*.6) circle[radius=0.02];
\filldraw (180-22.5-30:\R*.6) circle[radius=0.02];
\filldraw (180-17.5:\R*1.2) circle[radius=0.02];
\filldraw (180-17.5-10:\R*1.2) circle[radius=0.02];
\filldraw (180-17.5-20:\R*1.2) circle[radius=0.02];
\filldraw[fill opacity=0.5,fill=green] (45:\R) arc (45:-45:\R);
\fill[fill opacity=0.5, fill=green] (0,0)--(45:\R)--(-45:\R);
\end{tikzpicture}
\caption{Pictorial representation of rank-$1$ matrix factorizations (\ref{eq:startmf}).
Disk segments labelled by $[id']_d$ correspond to linear matrix factorizations, i.e.~$r=1$ with the respective charge label $[a]_d=[id']_d$. Unions of $r$ neighboring segments correspond to matrix factorizations with $r>1$. For instance the union of the two green segments above corresponds to the matrix factorization \ref{eq:startmf} with $r=2$ and $[a]_d=[d']_d$.
\label{fig:Atype}}
\end{figure}

In this way, one can explain the action of $RG_0$ on any rank-$1$ matrix factorization
\begin{equation}\label{eq:pgen}
P: S_{(Y,\cdot)}\{[-b+5r]_8\}
\\ 
	\tikz[baseline=0]{
		\node at (0,.5) {$Y^r$};
		\draw[arrow position = 1] (-0.3,.2) -- (0.3,.2);
		\draw[arrow position = 1] (0.3,0) -- (-0.3,0);
		\node at (0,-.4) {$Y^{8-r}$};
	}\, 
	S_{(Y,\cdot)}\{[-b]_8\}\,.
\end{equation}
The result can be read off from the general formulas above. We summarize it in the following table:
\begin{equation}\nonumber
\begin{array}{c|c||c||c}
\text{degree of }P: r& \text{charge shift of } P\text{ specified by }b &\text{charges of lift }T_N\otimes P&\text{degree of }P'': n \\
\hline
1 & 5\leq b <8 &-b, -b+5 & 0\\
1 & 0\leq b <5 & -b, -b-3&1\\
2 & 2 \leq b <8 & -b, -b+2&1\\
2 & 0 \leq b <2 & -b, -b-6&2\\
3 & 7 \leq b <8 & -b, -b+7&1\\
3 & 0 \leq b <7 & -b, -b-1&2 \\
4 & 4 \leq b <8 & -b, -b+4&2\\
4 & 0\leq b<4 & -b, -b-4&3\\
5 & 1\leq b <8 &-b, -b+1 &3\\
5 & 0\leq b <1 & -b, -b-7&4\\
6 & 6\leq b <8 & -b, -b+6&3\\
6 & 0\leq b<6 & -b, -b-2&4\\
7 & 3\leq b<8 & -b, -b+3&4\\
7 & 0\leq b<3 & -b, -b-5&5
\end{array}
\end{equation}
Indeed, there is a nice pictorial representation of the action of $RG_N$, which has its origin in the mirror dual of the branes \cite{Hori:2000ck} and  flows described by the GLSM. Consider a disk partitioned into $d$ segments labelled (clockwise) by the rest classes $[0]_d,[d']_d,[2d']_d,\ldots,[(d-1)d']_d$, c.f.~figure~\ref{fig:Atype}. Each of these corresponds to a linear matrix factorization, namely the matrix factorization (\ref{eq:p1}) with $r=1$ and $[a]_d$ given by the respective label of the segment. Now, the linear matrix factorizations corresponding to consecutive segements form non-trivial cones. More precisely, the matrix factorization (\ref{eq:startmf}) with $r>1$ can be obtained as successive cone of $r$ linear matrix factorizations labelled by $[a]_d,[a+d']_d,\ldots,[a+rd']_d$. In the picture we represent it by the union of the respective segments. The pictorial representation of matrix factorizations of the IR theory consists of a disk partitioned into $d'$ segments.

Under the action of $RG_N$ all linear factorizations labelled by $[a]_d$ with $\{N-a\}_d\geq d'$ are mapped to trivial matrix factorizations, while all the other ones are mapped to linear matrix factorizations of the IR theory, labelled by
$[N-\{N-a\}_d]_{d'}$. Note that linear factorizations in the UV theory labelled by $[a]_d$ and $[a+d']_d$, which are mapped to linear matrix factorizations under the action of $RG_N$, and which can form bound states (and are therefore represented by neighboring disk segments) are mapped to linear matrix factorizations in the IR theory, which can form bound states as well (and are therefore represented by consecutive segements in the pictorial representation of matrix factorizations of the IR theory).\footnote{This follows from the fact that
$\{N-a-d'\}_d-\{N-a\}_d$ is either $-d'$ if $\{N-a\}_d\geq d'$ or it is $d-d'$ in case $\{N-a\}_d<d'$.
Hence, 
$$
[N-\{N-a\}_d]_{d'}-[N-\{N-a-d'\}_d]_{d'}=\left\{\begin{array}{ll}
[d]_{d'}\,,&\{N-a\}_d<d'\\{}[0]_{d'}\,,&\{N-a\}_d\geq g'
\end{array}\right.\,.
$$
In the first case, both linear matrix factorization are mapped to linear matrix factorizations under $RG_N$, which can form a cone. In the second case the first one is mapped to a trivial matrix factorization.}
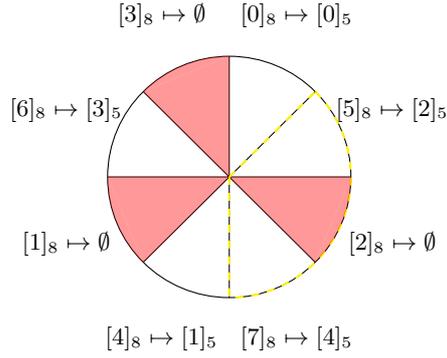
\begin{figure}[h] 
\centering
 \begin{tikzpicture}[scale=0.8]
\def\R{2cm}
\draw (0,0) circle[radius=\R];
\draw (0,0) -- (0:\R)    (0,0) -- (45:\R)     (0,0) -- (90:\R)   (0,0)--(135:\R)   (0,0) -- (180:\R)    (0,0)--(225:\R)    (0,0) -- (315:\R) (0,0)--(270:\R);
\draw[yellow, thick, dashed](0,0)--(45:\R) (0,0)--(-90:\R) (45:\R) arc (45:-90:\R);
\node[font=\footnotesize] at (90-22.5:\R*1.45) {$[0]_8\mapsto[0]_5$};
\node[font=\footnotesize] at ( 22.5:\R*1.45) {$[5]_8\mapsto[2]_5$};
\node[font=\footnotesize] at ( -22.5:\R*1.45) {$[2]_8\mapsto\emptyset$};
\node[font=\footnotesize] at ( -90+22.5:\R*1.45) {$[7]_8\mapsto[4]_5$};
\node[font=\footnotesize] at ( -135+22.5:\R*1.45) {$[4]_8\mapsto[1]_5$};
\node[font=\footnotesize] at ( -180+22.5:\R*1.45) {$[1]_8\mapsto\emptyset$};
\node[font=\footnotesize] at ( -225+22.5:\R*1.45) {$[6]_8\mapsto[3]_5$};
\node[font=\footnotesize] at ( -270+22.5:\R*1.45) {$[3]_8\mapsto\emptyset$};
\fill[fill opacity=0.4,fill=red] (180:\R) arc (180:225:\R);
\fill[fill opacity=0.4, fill=red] (0,0)--(180:\R)--(225:\R);
\fill[fill opacity=0.4,fill=red] (0:\R) arc (0:-45:\R);
\fill[fill opacity=0.4, fill=red] (0,0)--(0:\R)--(-45:\R);
\fill[fill opacity=0.4,fill=red] (90:\R) arc (90:135:\R);
\fill[fill opacity=0.4, fill=red] (0,0)--(90:\R)--(135:\R);
\end{tikzpicture}
\caption{Pictorial representation of action of $RG_0$ in the example of $d=8$, $d'=5$. The red segments correspond to linear matrix factorizations which are mapped to the trivial matrix factorization under $RG_0$. They are collapsed under the action. The arrows `$\mapsto$' indicate to which linear matrix factorizations in the IR the respective UV linear matrix factorization is mapped.
For instance $[0]_8\mapsto[0]_{5}$ means that the linear factorization with label $[0]_8$ in the UV is mapped to the linear factorization $[0]_{5}$ in the IR. From this, one can read off, how any rank-$1$-matrix factorization is mapped. For instance the matrix factorization with $r=3$ and $[a]_8=[5]_8$ corresponds to the union of consecutive segments $[5]_8$, $[2]_8$ and $[7]_8$, marked with yellow dashed boarder in the figure. Under the flow, segment $[2]_8$ is collapsed, and the other two obtain labels $[2]_5$ and $[4]_5$. They are neighboring in the IR, and their union corresponds to the matrix factorization in the IR with $r=2$ and $[a]_5=[2]_5$.
\label{fig:Atypeflow}}
\end{figure}

Thus, one arrives at the following pictorial representation of the action of $RG_N$, c.f.~figure~\ref{fig:Atypeflow}. $RG_N$ acts by shrinking the disk segments corresponding to decoupling linear matrix factorizations. These are the segments labelled by 
$[a]_d$ with $\{N-a\}_d\geq d'$. In this way one arrives at a disk partitioned into $d'$ segments. This is a pictorial representation of the matrix factorizations describing D-branes in the IR theory.  A segment labelled by $[a]_d$ in the original UV picture gets label $[N-\{N-a\}_d]_{d'}$ in the IR. One can now read off, what happens to a general matrix factorization under the action of $RG_N$ from this pictorial representation. A matrix factorization (\ref{eq:startmf}) corresponds to a union of $r$ consecutive segments in the UV picture. After shrinking the respective segments, it is represented by the union of $n<r$ consecutive segments in the IR picture, which is the representation of the matrix factorization (\ref{eq:p''}). For more details see \cite{Brunner:2007ur}.

\subsection{Comparison with other approaches}
D-Brane transport between phases of abelian gauged linear sigma models has been investigated before with very different methods. The non-anomalous ``Calabi-Yau'' case was studied in \cite{Herbst:2008jq}. A discussion going beyond abelian gauge groups as well as an extension to
anomalous models can be found in the more recent work \cite{Hori:2013ika,Clingempeel:2018iub}.

In \cite{Hori:2013ika,Clingempeel:2018iub}, hemisphere partition functions are computed 
in curved backgrounds with B-type boundary conditions on the equator by means of path integral localization. As a result of the curvature of the background, these precisely capture the dependence of B-type boundary conditions on the parameters appearing in the gauge sector. A 
thorough analysis of  analytic and convergence properties of hemisphere partition functions, then allows 
to determine the brane transport between different phases. This as well as the arguments in \cite{Herbst:2008jq} rely on a detailed analysis of the boundary conditions imposed in the gauge sector.

The approach taken in the present paper is very different. We decouple the gauge sector, and boundary conditions in this sector are not taken into account. Essentially\footnote{with the exception of the truncation, which we introduced to obtain the RG defects from the GLSM identity defect, and which presumably is related to stability}, we only consider information accessible to the B-twisted model. 
That means that we cannot control any analycity or explicit dependence on $t$. 
 Remarkably, our approach still yields many similar results that we highlight in the following.
 
 A crucial ingredient in the discussion of D-brane transport in \cite{Herbst:2008jq} as well as \cite{Hori:2013ika,Clingempeel:2018iub} are so called ``charge windows''. A D-brane whose $U(1)$-charges all lie in this window can be transported smoothly from one phase to another. Partition functions of these grade restricted branes are well behaved in both phases involved. Any D-brane in the GLSM has a grade restricted representative, which can be obtained by binding D-branes which are trivial in the phase in which the transport starts. 
The charge window is determined by the choice of the homotopy class of paths in parameter space, along which the D-branes are transported. 
 
In our approach, the defect $RG_N$ automatically takes care that branes are transported through such windows. Indeed the defect $T_N^{i}$ lifting a phase $i$ to the GLSM automatically maps D-branes from phase$_i$ to grade restricted GLSM branes, where the exact window is determined by the truncation parameter $N$. The projection defect $P_N^i$ realizing phase$_i$ in the GLSM projects the category of GLSM branes on the grade restricted subcategory, i.e. it maps every D-brane to the respective grade restricted representative.

Note that a in the treatment of \cite{Hori:2013ika,Clingempeel:2018iub} a D-brane transport between two phases actually involves two charge windows, a ``large window'' which ensures smooth transport as alluded to above, and a ``small window'' lying in the large one \footnote{The two windows coincide in the Calabi-Yau case.}. D-branes, whose charges completely lie in the small window flow to the new conformal fixed points, while D-branes, whose charges lie in the large window, but not completely in the small one undergo some kind of decay. (In \cite{Hori:2013ika,Clingempeel:2018iub} this is determined by analyzing the asympotics of the hemisphere partition functions.)

 In our approach both these windows appear naturally and on the same footing. The large window is determined by the projection $P_N^{\rm UV}$ associated to the phase, in which the transport starts, and the small window comes from the projection $P_N^{\rm IR}$ associated to the phase, in which the transport ends. Indeed, on the level of the GLSM the transport from phase $i$ to phase $j$ can be described by the fusion $P^{\rm IR}_N\otimes P^{\rm UV}_N$ of the respective projection defects.

Transporting  branes from one phase to another can involve monodromies.  In \cite{Hori:2013ika,Clingempeel:2018iub} these are naturally associated with shifts in the two windows, either the large window as a whole, or the small window inside the large window. In our case, the windows are determined by the truncation parameters $N$, which can be shifted
by a quantum symmetry, which exactly realizes the monodromy around the fixpoint of the respective phase.

Transporting branes from the GLSM to a phase can be done using two different functors. The authors of \cite{Clingempeel:2018iub} consider geometric phases and define two functors $F_{{\rm flow}}$ and $F_{{\rm geom}}$. The first one corresponds to the actual flow from the GLSM to the phase, the second one to a restriction to field configuration allowed by the deleted sets of the toric geometry/GLSM description. In our case, we have two defects from the GLSM to a given phase$_i$, $R_N^{i}$ and $R^{i}$, the truncated and the untruncated descent defects.
$R_N^{i}$ depends on the the truncation parameter, and hence a path in parameter space, whereas $R^{i}$ merely sets certain fields to $1$. So these are precisely the analogues of $F_{{\rm flow}}$ and $F_{{\rm geom}}$. In the same way as in  \cite{Clingempeel:2018iub}, where the two functors agree on grade restricted branes, we have
$R^j_N\otimes P^i_N\cong R^j\otimes P^i$.
This is also the reason, why $R^{{\rm IR}}_N$ does not feature more prominently in our discussion: The lifts $T_N^{{\rm UV}}$ directly lift the UV phase to  grade restricted branes, and we chose to factorize $RG_N=R^{{\rm IR}} \otimes T_N^{{\rm UV}}$. We could have used the cutoff version of $R^{\rm IR}$ as well, writing equivalently $RG_N=R^{{\rm IR}}_N \otimes T_N^{{\rm UV}}$.
 
\medskip

One reason, why our approach, which is essentially based on the B-twisted model, still captures all this information might be the fact that functoriality is a strong constraint. Functoriality is inherent in the defect approach, and B-type defects seem to be rather rigid. With the exception of the truncation, which we introduced in an ad-hoc fashion to obtain RG defects from the GLSM identity defect, and which probably has its origins in stability considerations, there were no choices involved in our construction. Furthermore, this choice exactly aligns with the choice of paths between the respective phases.

It would be very interesting, to understand the relation of our approach to the ones in \cite{Herbst:2008jq,Hori:2013ika,Clingempeel:2018iub} even better. For one thing, in \cite{Hori:2013ika,Knapp:2020oba}
the D-brane central charge and concrete dependence on the twisted chiral moduli is investigated quite explicitly. In particular, in \cite{Knapp:2020oba}  the mathematics of central charges in Landau-Ginzburg orbifolds is studied in detail. By general arguments, we expect that RG (or deformation) defects act on these objects via fusion, and it should be possible to formulate this operation in a natural way. On the other hand, one could try to incorporate the functoriality
constraint directly into the approach of \cite{Hori:2013ika,Clingempeel:2018iub} by applying their analysis to the
GLSM identity defect constructed in section~\ref{sec:identity}.

\section{Conclusions}

In this paper, we have constructed defects that concretely describe the behavior of  D-branes 
under transitions between phases of abelian gauged linear sigma models.
They act on objects and morphisms of the respective D-brane categories via fusion, and this action is automatically functorial. A key ingredient is the new construction of the identity defect in gauged linear sigma models presented in section~\ref{sec:identity}. Our approach gives a novel perspective on earlier work \cite{Herbst:2008jq,Hori:2013ika,Clingempeel:2018iub} on D-brane transport in GLSMs.
We conclude this paper with a list of interesting points for future investigation.
\begin{itemize}
\item
The starting point for the construction of our defects $RG_N$ that implement the transition between a UV and IR Landau-Ginzburg phase of a $U(1)$-gauged linear sigma model is the identity defect of the GLSM. The bosonic defect fields that we use to construct it create an infinite dimensional Chan-Paton-like space. In other words, the modules on which the associated equivariant matrix factorization is built are of infinite rank. Introducing a finite cutoff $N$ for these modules, we obtain defects $RG_N$ in agreement with expectations and earlier results \cite{Brunner:2007ur}.
The choice of cutoff corresponds to a choice of homotopy class of paths in K\"ahler parameter space. 
While we formulate all defects and boundaries on the level of the B-twisted model, which decouples from K\"ahler parameters, a (mild) K\"ahler dependence sneaks back in via the cutoff. We expect the choice of cutoff to be related to stability, one of the indicators being that the cutoff is necessary to ensure consistent gluing conditions on a spectral flow operator of an IR conformal field theory. It would be very interesting to investigate further, whether stability conditions in phases can be discussed on the level of the GLSM, and how this relates to defects.
\item
It would be very interesting to combine our approach with the one of 
\cite{Herbst:2008jq,Hori:2013ika,Clingempeel:2018iub}. Applying their methods to the GLSM identity defect would 
at the same time 
explicitly incorporate the constraint of functoriality in their approach as well as elucidate the precise origin of the cutoff appearing in our construction.
\item
In section~\ref{sec:example} we applied the general approach outlined in section~\ref{sec:transition} to a specific class of $U(1)$-gauged linear sigma models which only exhibit Landau-Ginzburg phases. It would be very interesting to apply it to more interesting models, in particular those featuring geometric or mixed phases. Indeed, a paper, in which we employ our methods to models with geometric phases is already in preparation \cite{forthcoming}.
\item
The construction of the identity defect should also generalize to non-abelian gauged linear sigma models. 
It would be very interesting to spell this out and obtain transition and monodromy defects also for phases of non-abelian GLSMs.
\item
While in two dimensions our methods are particularly powerful, as the fusion of defects is well-controllable, our basic ideas are not limited to this and it would be quite interesting to discuss phase transitions and possibly dualities from this point of view also in higher dimensions. 
\end{itemize}

\section*{Acknowledgements}

IB thanks Lukas Krumpeck for discussions and Christoph G\"artlein for comments on the manuscript.
FK is thankful to Friedrich-Naumann-Stiftung for supporting this project. DR is supported by the Heidelberg Institute for Theoretical Studies. DR also thanks the MSRI in Berkeley for its hospitality, where part of this work was done. (Research at MSRI is partly supported by the NSF under Grant No. DMS-1440140.) IB is supported by the Excellence Cluster Origins and the DFG.

\appendix
\section{Defects in LG models and their orbifolds}

In this appendix, we summarize some aspects of the description of defects in Landau-Ginzburg models and their orbifolds  in terms of (equivariant) matrix factorizations, paying particular attention to defect fusion. The exposition is very brief. For a more detailed exposition we refer to the literature. In particular, matrix factorizations where related to categories of D-branes in \cite{Orlov:2003yp} from a mathematical point of view, and in \cite{Kapustin:2002bi,Brunner:2003dc,Lazaroiu:2003zi} from a physical point of view. We refer to \cite{Hori:2004zd} for a brief summary of the basic physical aspects.
The description of defects in Landau-Ginzburg models by means of matrix factorizations has been established in \cite{Khovanov:2004bc,Brunner:2007qu,Brunner:2007ur}, see
\cite{Carqueville:2012st} for an exposition emphasizing categorical aspects.

\subparagraph{Defects  in LG models.} 
 A defect $D:W\rightarrow V$ from a LG model with chiral fields $X_1,\dots X_n$ and superpotential $W\in\bC[X_1, ..., X_n]$ to a LG model specified by $V\in\bC[Z_1, ..., Z_m]$ is described by a matrix factorization of the difference of the potentials $V-W$ over $S:=\bC[Z_1, ..., Z_m, X_1, ..., X_n]$. That is, there is a $\bZ_2$-graded free module $D=D_0\oplus D_1$ over $S$ with an odd endomorphism
\[
	\d_D=\begin{pmatrix}0 & \d_{D1} \\ \d_{D0} & 0\end{pmatrix}\quad\text{such that }\d^2 = W\cdot \id_D.
\]
Matrix factorizations can also be regarded as two-periodic complexes twisted by $W$
\[
	D:D_1\,
	\tikz[baseline=0]{
		\node at (0,.6) {$\d_{D1}$};
		\draw[arrow position = 1] (-.8,.2) -- (.8,.2);
		\draw[arrow position = 1] (.8,0) -- (-.8,0);
		\node at (0,-.4) {$\d_{D0}$};
	}\,D_0
\]
with $\d_{D1}\cdot\d_{D0} = W\cdot\id_{D0}$ and $\d_{D0}\cdot\d_{D1} = W\cdot\id_{D1}$.\footnote{As is customary in the literature, $D$ refers to the free module as well as the defect and matrix factorization.}

(Right) boundary conditions are the special class of defects, for which the right LG model is trivial, i.e.~does not feature any chiral fields. 

To a matrix factorization $D$ as above one can associate an $R=S/(W)$-module 
$$
M=\text{coker}(\d_{D1}:D_1\otimes_S R\rightarrow D_0\otimes_S R)\,,
$$
which has a two-periodic free resolution induced by the matrix factorization
$$
 \dots  \xlongrightarrow{{\d_{D1}}}
D_0\otimes_S R \dots  \xlongrightarrow{{\d_{D0}}} D_1\otimes_S R \xlongrightarrow{{\d_{D1}}} D_0\otimes_S R \longrightarrow M \longrightarrow 0 \ .
$$
More generally, it is useful to consider $R$-modules which have free resolutions which, after finitely many steps turn into the resolutions induced by the matrix factorizations.
Such $R$-modules can be used for instance to find isomorphisms between different matrix factorizations. This is explained in section~\ref{sec:mfmodules} and we also refer to \cite{Enger:2005jk} for an exposition that is useful for the line of arguments in the current paper.

The fusion of defects is given in terms of the tensor product of matrix factorizations  \cite{Brunner:2007qu}. More precisely, consider superpotentials $U\in\mathbb{C}[X_1,\ldots,X_m]$, $V\in\mathbb{C}[Y_1,\ldots,Y_n]$, $W\in\mathbb{C}[Z_1,\ldots,Z_o]$ together with  matrix factorizations $D$ of $V-W$, and  $D'$ of $U-V$. The tensor product $D'\otimes D$ is then a matrix factorization of $U-W$, with modules
\begin{equation}
(D'\otimes D)_0= D'_0\otimes_{{\mathbb C}[Y_i]} D_0 \oplus D'_1 \otimes_{{\mathbb C}[Y_i]} D_1, \quad (D'\otimes D)_1= D'_1\otimes_{{\mathbb C}[Y_i]} D_0 \oplus D'_0 \otimes_{{\mathbb C}[Y_i]} D_1
\end{equation}
and differential
\begin{equation}\label{eq:defect fusion}
		\d_{D' \otimes D} = \d_{D'} \otimes \text{id}_D + \text{id}_{D'} \otimes \d_D\,.
\end{equation}
	This differential is to be understood with Koszul signs, meaning that
\[
		(\text{id}_{D'}\otimes \d_{D}) (\nu \otimes \omega) = (-1)^{\text{deg}}(\nu) \otimes \d_{D}(\omega).
\]
While this tensor product a priori has infinite rank over ${\mathbb C}[X_i, Z_j]$ it can be shown that it is always isomorphic to a finite rank matrix factorization. 

\subparagraph{Defects  in LG orbifolds.} A symmetry of a Landau-Ginzburg model is a homomorphism of the ring of chiral fields, which leaves the superpotential invariant. Given a group of symmetries of a Landau-Ginzburg model, one can take the orbifold by that group. 
Defects and boundaries in Landau-Ginzburg orbifolds are well understood in the case of finite orbifold groups. 

Let $V\in\bC[X_1,\ldots,X_n]$ and $W\in\bC[Y_1,\ldots,Y_m]$ be two superpotentials and $G_V$ and $G_W$  orbifold groups of the respective LG models. 
Then B-type defects between these models can be described by   $G=G_V\times G_W$-equivariant matrix factorizations of $V-W$ \cite{Brunner:2007ur,Ashok:2004zb,Hori:2004ja}. The defects are equipped with a representation $\rho_D$ of $G$ that is compatible with the action of the combined polynomial ring $S:=\bC[X_1,\ldots,X_n,Y_1,\ldots,Y_m]$ on the modules $D_0$, $D_1$.   Denoting by $\rho$ the representation of $G=G_V\times G_W$ on $S$ this means that for all $g\in G$
\begin{eqnarray*}
&&\rho_D(g)(s\cdot p) = \rho(g)(s) \cdot \rho_D(g)(p), \qquad \forall s\in S, p\in D=D_0\oplus D_1,\\
&&\rho_{D}(g)\circ \d_{D} = \d_{D} \circ \rho_D(g)\,.\nonumber
\end{eqnarray*}
We are interested in abelian orbifold groups here. The action of the latter 
gives the polynomial ring $S$ the structure of a graded ring, and the representations $\rho_D$ turn $D_1$ and $D_0$ into graded $S$-modules. The map $\d_D$ respects the grading and one can associate grade $0$ to it. In all our discussions in this paper, the group action can be diagonalized on the modules, and we will often pick 
generators of $D_0$ and $D_1$ on which $\rho_D$ acts diagonal. The action of $G=G_V \times G_W$ can then be specified by assigning
$G_V\times G_W$ grades to the generators. We denote these grades (or rather their shifts) using curly brackets. To illustrate our notation, we would specify a rank $k$ ${\mathbb Z}_d \times {\mathbb Z}_{d'}$ equivariant matrix factorization as follows
\begin{equation*}
	D: S^k\begin{pmatrix}\{[l_k]_d,[r_k]_{d'}\}\\ \{[l_{k+1}]_d,[r_{k+1}]_{d'}\}\\ \{[l_{k+2}]_d,[r_{k+2}]_{d'}\} \\ \vdots \end{pmatrix}
	\tikz[baseline=0]{
		\node at (0,.5) {$\d_{D1}$};
		\draw[arrow position = 1] (-2.5,.2) -- (2.5,.2);
		\draw[arrow position = 1] (2.5,0) -- (-2.5,0);
		\node at (0,-.4) {$\d_{D0}$};
	}\,S^k\begin{pmatrix}\{[l_0]_{d},[r_0]_{d'}\}\\ \{[l_1]_d,[r_1]_{d'}\}\\ \{[l_2]_d,[r_2]_{d'}\} \\ \vdots \end{pmatrix} \ .
\end{equation*}
Here  $l_i$ and $r_i$ are integers, $[ \ . \ ]_d$ denotes the rest class  modulo $d$, and $\{[l_i]_d,[r_i]_{d'}\}$ signify that the respective generator in $S^k$ carries $\bZ_d\times\bZ_{d'}$-charge $([l_i]_d,[r_i]_{d'})$. 

Defect fusion can be extended to orbifold LG theories in a straight-forward way \cite{Brunner:2007ur}. Let $U\in\mathbb{C}[X_1,\ldots,X_m]$, $V\in\mathbb{C}[Y_1,\ldots,Y_n]$ and $W\in\mathbb{C}[Z_1,\ldots,Z_o]$ be polynomials invariant under actions of groups $G_U,G_V,G_W$ on the respective polynomial rings. 
Consider a $G_W\times G_V$ equivariant matrix factorization $D$ of $U-V$ and a $G_V\times G_U$-equivariant matrix factorizations $E$ of $V-W$. The tensor product $D\otimes E$ a priori yields a 
 $G_U\times G_V\times G_W$-equivariant matrix factorization of $U-W$. 
The fusion in the orbifold theory is now given by the $G_V$ invariant part of $(D\otimes E)^{G_V}$ of the tensor product factorization $D\otimes E$. This yields a
$G_U\times G_W$ equivariant factorization of $U-W$ over ${\mathbb C}[X_1, \dots, X_m, Z_1, \dots Z_o]$.

	\bibliographystyle{JHEP}
	\bibliography{references}

\end{document}